\documentclass{article} % For LaTeX2e
\usepackage{arxiv_iclr2026_conference,times}

\usepackage{hyperref}
\usepackage{url}

\usepackage{booktabs}       % professional-quality tables
\usepackage{amsfonts}       % blackboard math symbols
\usepackage{nicefrac}       % compact symbols for 1/2, etc.
\usepackage{microtype}      % microtypography
\usepackage{xcolor}         % colors

\usepackage{amsmath}
\usepackage{graphicx} % Required for inserting images
\usepackage[]{algpseudocode}

\usepackage{algorithm}

\usepackage{mathrsfs}
\usepackage{bbding} % \Envelope
\usepackage{diagbox}
\usepackage{rotating}

\newcommand{\blue}[1]{{#1}}

\title{Monotone Near-Zero-Sum Games:\\ A Generalization of Convex-Concave Minimax}

\author{Ruichen Luo\thanks{This work was partially done during RL's stay at CISPA. } \\
  IST Austria\\
  3400 Klosterneuburg, Austria \\
  \And
  Sebastian U. Stich \\
  CISPA Helmholtz Center \\
  66386 St. Ingbert, Germany \\
  \And
  Krishnendu Chatterjee \\
  IST Austria \\
  3400 Klosterneuburg, Austria \\
}

\iclrfinalcopy % Uncomment for camera-ready version, but NOT for submission.

\usepackage{amssymb,amsmath,amsthm,dsfont}
\usepackage{bm}
\usepackage{mdframed}

\providecommand{\lin}[1]{\ensuremath{\left\langle #1 \right\rangle}}
\providecommand{\abs}[1]{{\textnormal{abs}}\left(#1\right)}
\providecommand{\norm}[1]{\left\lVert#1\right\rVert}

\providecommand{\defEQ}{\stackrel{\textnormal{def}}{=}}

\providecommand{\refLE}[1]{\ensuremath{\stackrel{(\ref{#1})}{\leq}}}

  \providecommand{\R}{\mathbb{R}} % Reals
  
  \DeclareMathOperator*{\argmin}{arg\,min}
  \DeclareMathOperator*{\argmax}{arg\,max}

  \providecommand{\1}{\mathbf{1}}
  \renewcommand{\aa}{\mathbf{a}}
  \providecommand{\bb}{\mathbf{b}}

  \let\lll\ll
  \renewcommand{\ll}{\mathbf{l}}

  \providecommand{\uu}{\mathbf{u}}
  \providecommand{\vv}{\mathbf{v}}
  
  \providecommand{\xx}{\mathbf{x}}
  \providecommand{\yy}{\mathbf{y}}
  \providecommand{\zz}{\mathbf{z}}
  
  \providecommand{\mA}{\mathbf{A}}
  \providecommand{\mB}{\mathbf{B}}

  \providecommand{\mM}{\mathbf{M}}

  \providecommand{\cF}{\mathcal{F}}
  
  \providecommand{\cH}{\mathcal{H}}

  \providecommand{\cO}{\mathcal{O}}
  \providecommand{\cP}{\mathcal{P}}
  \providecommand{\cQ}{\mathcal{Q}}
  \providecommand{\cR}{\mathcal{R}}

  \providecommand{\cX}{\mathcal{X}}
  \providecommand{\cY}{\mathcal{Y}}

\RequirePackage[colorinlistoftodos,bordercolor=orange,backgroundcolor=orange!20,linecolor=orange,textsize=scriptsize]{todonotes}

\definecolor{mydarkblue}{rgb}{0,0.08,0.45}
\usepackage{hyperref} 
\usepackage{thmtools}
\usepackage[capitalize,noabbrev]{cleveref} % must be loaded after hyperref
\crefname{relctr}{relation}{Relation} %% <- if you use cleveref

\newtheorem{lemma}{Lemma}
\newtheorem{proposition}[lemma]{Proposition}
\newtheorem{corollary}[lemma]{Corollary}
\newtheorem*{definition}{Definition}

\newtheorem{assumption}{Assumption}
\crefname{assumption}{Assumption}{Assumptions}
\newtheorem{theorem}{Theorem}

\crefalias{strongerassumption}{assumption}

\declaretheoremstyle[
  spaceabove=\topsep, spacebelow=\topsep,
  headfont=\normalfont\scshape, 
  notefont=\normalfont, notebraces={(}{)},
  bodyfont=\normalfont\itshape,
  postheadspace=1em
]{mythmstyle}
\declaretheorem[style=mythmstyle]{remark}
\declaretheorem[style=mythmstyle]{example}

\usepackage{url}

\usepackage{breakcites}
\usepackage{nicefrac}
\usepackage[normalem]{ulem}

\usepackage{enumitem}% http://ctan.org/pkg/enumitem
\setlist[itemize]{leftmargin=12pt, itemsep=0pt, topsep=0.5pt}
\setlist[enumerate]{leftmargin=12pt, itemsep=0pt, topsep=0.5pt}

\renewcommand{\tilde}{\widetilde} % make \tilde wilder

\renewcommand{\epsilon}{\varepsilon}

\renewcommand{\phi}{\varphi}

\renewcommand{\Gamma}{\varGamma}

\renewcommand{\Delta}{\varDelta}

\renewcommand{\Theta}{\varTheta}

\renewcommand{\Lambda}{\varLambda}

\renewcommand{\Xi}{\varXi}
\let\oldPi\Pi
\renewcommand{\Pi}{\varPi}

\renewcommand{\Sigma}{\varSigma}

\renewcommand{\Upsilon}{\varUpsilon}

\renewcommand{\Phi}{\varPhi}

\renewcommand{\Psi}{\varPsi}

\renewcommand{\Omega}{\varOmega}

\usepackage{xspace}
\usepackage{bm}
\newcommand{\algname}[1]{{\sf\footnotesize#1}\xspace}

\usepackage{pifont}
\usepackage{etoolbox}
\newcounter{relctr} %% <- counter for relations
\everydisplay\expandafter{\the\everydisplay\setcounter{relctr}{0}} %% <- switch to the following better alternative
\makeatletter
  \def\blackcirclenumbers#1{\expandafter\@blackcirclenumbers\csname c@#1\endcsname}
  \def\@blackcirclenumbers#1{%
    \ifcase#1\or \ding{182}\or \ding{183}\or \ding{184}\or \ding{185}\or \ding{186}\or \ding{187}\or \ding{188}\or \ding{189}\or \ding{190}\or \ding{191}\or \else\@ctrerr\fi}
\makeatother
\AtBeginDocument{} %% <- store original definition
\usepackage{tcolorbox}
\usepackage{pifont}
\definecolor{mydarkgreen}{RGB}{39,130,67}
\definecolor{mydarkred}{RGB}{192,25,25}

\providecommand{\pone}{{Player~1}}
\providecommand{\ptwo}{{Player~2}}
\providecommand{\fone}{{u_1}}
\providecommand{\ftwo}{{u_2}}

\providecommand{\xnorm}[1]{\norm {#1}}

\providecommand{\znorm}[1]{\norm {#1}}

\providecommand{\dznorm}[1]{\norm {#1}}
\providecommand{\mnorm}[1]{\norm {#1}}
\providecommand{\ICL}{\algname{Iterative Coupling Linearization}}
\providecommand{\com}{\textnormal{b}}
\providecommand{\coo}{\textnormal{a}}

\DeclareMathOperator*{\dom}{dom}
\DeclareMathOperator{\prox}{prox}

\renewcommand{\paragraph}[1]{\noindent\textbf{#1 }}
\begin{document}

\maketitle

\begin{abstract}
Zero-sum and non-zero-sum (aka general-sum) games are relevant in a wide range of applications. 
While general non-zero-sum games are computationally hard, researchers focus on the special class of monotone games for gradient-based algorithms.
However, there is a substantial gap between the gradient complexity of monotone zero-sum and monotone general-sum games. 
Moreover, in many practical scenarios of games the zero-sum assumption needs to be relaxed.
To address these issues, we define a new intermediate class of \emph{monotone near-zero-sum games} that contains monotone zero-sum games as a special case. Then, we present a novel algorithm that transforms the near-zero-sum games into a sequence of zero-sum subproblems, improving the gradient-based complexity for the class. Finally, we demonstrate the applicability of this new
class to model practical scenarios of games motivated from the literature. 
\end{abstract}

\section{Introduction}

Two-player zero-sum games (also known as strictly competitive games) and their generalization to non-zero-sum games~\citep{nash1951non,rosen1965existence} are crucial in domains like economics~\citep{von1947theory}, artificial intelligence~\citep{yannakakis2018artificial}, and biology in the form of evolutionary game theory~\citep{weibull1997evolutionary,smith1982evolution}. 
As initiated in \cite{rosen1965existence,tseng1995linear,nemirovski2004prox}, this paper focues on the computationally tractable class of \emph{monotone games} {with compact convex strategy spaces}. 

% \footnote{The assumption of compact convex strategy spaces are quite common in the study of games. Theoretically, von Neumann's and Sion's minimax theorem~\citep{von1928minimax, sion1958minimax} both require the strategy spaces to be compact and convex. Practically, for games with discrete action sets, we may consider probability distributions over the action sets and take the expectation of the players. For instance, in classic games like rock-paper-scissors, by considering probability distributions over actions, the unique Nash equilibrium is for both players to play rock/paper/scissors with probability $\nicefrac{1}{3}$~\citep{nash1951non}.} 

Early work studies the restrictive classes of games where the two players have \emph{equal conditioning}: the seminal paper of \cite{nesterov2005smooth} established the complexity for bilinearly-coupled zero-sum games, which was later extended by \cite{nemirovski2004prox} to general-sum classes. Later on, the field explores the broader concept of \emph{general conditioning}, which allows for more detailed portrayal of the two players~\citep{chambolle2011first,lin2020near}. Early explorations of \cite{chambolle2011first,chen2014optimal} focus on the bilinearly-coupled cases; starting from \cite{lin2020near}, there has been extensive research into the generally-coupled, generally-conditioned, and zero-sum games~\citep{yang2020catalyst,wang2020improved,kovalev2022first,lan2023novel,boct2023accelerated}. However, it has been an interesting open question whether these recent developments can be further generalized to the broader classes of generally-coupled, generally-conditioned, and \emph{non-zero-sum} games. 

On the theory side, there is a substantial gap between the gradient complexity of monotone zero-sum games and that of monotone general-sum games. The recent developments in minimax optimization~\citep{lin2020near,kovalev2022first,lan2023novel} establish much better complexity results for the zero-sum cases compared to the long-standing bounds of general-sum games~\citep{rockafellar1976monotone,tseng1995linear}. 

On the application side, strictly competitive scenarios, modeled by monotone zero-sum games, are often insufficient. Real-world game settings frequently involve factors such as transaction fees or semi-cooperation~\citep{kalai2013cooperation,halpern2013cooperative}, necessitating a relaxation of the zero-sum assumption in the modeling. 

\paragraph{Our contributions}
Our work makes the first step towards bridging the gap between the monotone zero-sum and general-sum classes. 
For this purpose, we introduce a new {intermediate} class of monotone games, present a novel algorithm for this class, and show the applicability of this new class. In detail: 

\begin{itemize}
    \item \textit{Theoretical motivation.} Our work extends the recent studies of monotone zero-sum games to the more general non-zero-sum settings. Specifically, we define a new intermediate class of games called \emph{monotone near-zero-sum games}, characterized by a smoothness parameter $\delta$ describing the game's proximity to a zero-sum game. This new class of games presents a natural interpolation between monotone zero-sum games and a class of monotone general-sum games based on the near-zero-sum parameter $\delta$, and thus, it partially bridges the gap of the monotone zero-sum and general-sum classes. 
    \item \textit{Main theoretical result.} We propose a novel algorithm, \emph{\ICL}~(\algname{ICL}), that provides a black-box reduction from monotone near-zero-sum games to zero-sum games. It converges to an $\epsilon$-Nash equilibrium within $\tilde \cO \left( \left(\frac {L} {\sqrt {\mu \nu}} + \frac {L} {\min \{\mu, \nu\}} \cdot \min \left\{ 1, \sqrt{ \frac {\delta} {\mu + \nu} } \right\} \right) \cdot \log ^2 \left( \frac {D^2} \epsilon \right) \right)$ gradient queries,\footnote{In the $\tilde \cO (\cdot)$ notations, the poly-logarithm terms are omitted.} where $L$ is the smoothness parameter, $\mu$ and $\nu$ are the strong concavity parameters of the two players, $\delta$ is the near-zero-sum parameter, and $D$ is the diameter. When $\delta$ is small, our results improve the long-standing complexity results of \cite{tseng1995linear,nemirovski2004prox} for the first time in non-zero-sum classes. 
    \item \textit{Practical applications.} Besides the theoretical motivation, we demonstrate the practical relevance of this new class of games. We consider \emph{regularized matrix games} and \emph{competitive games with small additional incentives}. These games are not zero-sum but naturally have a near-zero-sum structure, where our methods are applied to achieve provably faster rates. 
    % Even in the well-studied context of matrix games, this acceleration from the near-zero-sum structure \emph{and} general conditioning seems to be a new result, to the best of our knowledge. 
\end{itemize}

\section{Definitions, previous results, and the new problem class}

\subsection{Basic definitions}
This paper studies the \emph{Nash Equilibrium Problem} (NEP) for two-person general-sum games, in which \pone\ wants to maximize its utility function $u_1(\xx, \yy)$ over $\xx \in X$ and \ptwo\ wants to maximize its utility function $u_2(\xx, \yy)$ over $\yy \in Y$. Here, $X$ and $Y$ are compact and convex sets, and $u_1(\cdot, \cdot) \colon X \times Y \to \R$ and $u_2(\cdot, \cdot) \colon X \times Y \to \R$ are smooth functions. A pair of decisions $(\xx^*, \yy^*) \in X \times Y$ is a \emph{Nash equilibrium} if 
\[
u_1(\xx^*, \yy^*) \ge u_1(\xx, \yy^*) , \text{ for all } \xx \in X , \text{ and } u_2(\xx^*, \yy^*) \ge u_2(\xx^*, \yy) , \text{ for all } \yy \in Y \,. 
\]
A pair of decisions $(\bar \xx, \bar \yy) \in X \times Y$ is an $\epsilon$-\emph{accurate Nash equilibrium} if there exists a Nash equilibrium $(\xx^*, \yy^*)$ such that 
\(
\norm {\bar \xx - \xx^*} ^2 + \norm {\bar \yy - \yy^*} ^2 \le \epsilon \,. 
\)
A pair of decisions $(\hat \xx, \hat \yy) \in X \times Y$ is an $\epsilon$-\emph{approximate Nash equilibrium}, if 
\(
u_1 (\hat \xx, \hat \yy) \ge u_1 (\xx, \hat \yy) - \epsilon \text{ for all } \xx \in X 
\)
and 
\(
u_2 (\hat \xx, \hat \yy) \ge u_2 (\hat \xx, \yy) - \epsilon \text{ for all } \yy \in Y \,.
\)
The relation between accurate and approximate Nash equilibria is discussed in \cref{sec:different Nash equilibrium}. 
The goal of this paper is to find an $\epsilon$-accurate (or an $\epsilon$-approximate) Nash equilibrium by iterative algorithms which subsequently query the gradients of the utility functions. 
% To avoid ambiguity, when using the term \emph{gradient complexity} of an NEP, we are referring to the number of gradient queries needed to find an $\epsilon$-accurate Nash equilibrium. 

\paragraph{Notations}
Let $\cX$ and $\cY$ be Euclidean spaces. In the space $\cX \times \cY$, for all $\zz = (\xx, \yy) \in \cX \times \cY$ and $\zz^\prime = (\xx^\prime, \yy^\prime) \in \cX \times \cY$, define $\lin {\zz^\prime, \zz} \defEQ \lin {\xx^\prime, \xx} + \lin {\yy^\prime, \yy}$. For all these spaces, the norms are those induced by inner products. Assume that the diameter of $X \subseteq \cX$ is bounded by $D_X$ and the diameter of $Y \subseteq \cY$ is bounded by $D_Y$. Let $D = \sqrt{D_X^2 + D_Y^2}$. 
Assume that $\fone(\cdot, \cdot)$ and $\ftwo(\cdot, \cdot)$ are $L$-smooth, that is, 
\[
\dznorm { \nabla \fone (\zz ^\prime) - \nabla \fone (\zz) } \le L \znorm { \zz ^\prime - \zz }, \ \dznorm { \nabla \ftwo (\zz ^\prime) - \nabla \ftwo (\zz) } \le L \znorm { \zz ^\prime - \zz }, \text{ for all } \zz, \zz ^\prime \in X \times Y \,. 
\] 

To facilitate our analysis, we adopt the following formulation that decomposes the game into a coupling part and a zero-sum part. 
Denote 
\[
g = - \frac 1 2 (u_1 + u_2), \quad h = \frac 1 2 (- u_1 + u_2), \quad \cH = \left( \nabla _\xx h, -\nabla _\yy h \right), \quad \cF = - \left(
    \nabla _\xx u_1, \nabla _\yy u_2 
\right) \,, 
\]
where $g$ is the coupling part, $h$ is the zero-sum part, $\cH$ is the operator corresponding to the zero-sum part $h$, and $\cF$ is the operator corresponding to the game. Then, we have 
\[
u_1 = -g - h, \quad u_2 = -g + h, \quad \cF = \nabla g + \cH \,. 
\]
Since the utilities $u_1$ and $u_2$ are both $L$-smooth, we have the functions $g$ and $h$ are both $L$-smooth, and the operators $\cH$ and $\cF$ are both $L$-Lipschitz continuous. 
While similar decompositions can be found in the literature of variational inequalities and game theory~\citep{nemirovski1995information,halpern2013cooperative,chen2017accelerated,hwang2020strategic}, we emphasize that this notation is particularly suited to characterize our near-zero-sum games (to be defined later) for explicitly separating the non-zero-sum coupling part. 

\subsection{Problem classes}

% \paragraph{General-sum games and PPAD-hardness}
% The seminal work of \cite{rosen1965existence} establishes the existence of Nash equilibrium for concave games, where $\fone (\cdot, \yy)$ is concave for any fixed $\yy \in Y$ and $\ftwo (\xx,\cdot)$ is concave for any fixed $\xx \in X$. Therefore, in this context, computing a Nash equilibrium, or an $\epsilon$-accurate Nash equilibrium, is a well-defined problem. 
% However, without further restriction, the NEP for concave general-sum games is known to be PPAD-hard~\citep{chen2009settling,papadimitriou2023computational}. 

\paragraph{Monotone general-sum games}
The seminal work of \cite{nash1951non,rosen1965existence} establishes the existence of Nash equilibrium for concave games. 
But to obtain a tractable class of NEPs, further restrictions need to be considered~\citep{rosen1965existence}. Specifically, we make the following assumptions: 

\begin{assumption}[Convex-concave zero-sum part]
\label{assumption:convex-concave-combination}
    There exists $(\mu, \nu) \in [0, L] \times [0, L]$ such that the function $h(\xx,\yy) - \frac \mu 2 \norm {\xx}^2$ is convex in $\xx$ for any fixed $\yy \in Y$, and the function $h(\xx, \yy) + \frac \nu 2 \norm {\yy}^2$ is concave in $\yy$ for any fixed $\xx \in X$. 
\end{assumption}

\begin{assumption}[jointly convex coupling part]
\label{assumption:convex average}
    The function $g(\cdot,\cdot)$ is jointly convex. 
\end{assumption}

The operator $\cH = \left( \nabla _\xx h, -\nabla _\yy h \right)$ is monotone with modulus $\min \{ \mu, \nu \}$ under \cref{assumption:convex-concave-combination}, and the operator $\nabla g$ is monotone under \cref{assumption:convex average}. Hence, under \cref{assumption:convex-concave-combination,assumption:convex average}, the game (or the operator $\cF = \nabla g + \cH$) is monotone with modulus $\min \{ \mu, \nu \}$~\citep{rosen1965existence,nemirovski1995information}, that is, 
\(
\lin {\cF(\zz ^\prime) - \cF(\zz), \zz ^\prime - \zz} \ge \min \{ \mu, \nu\} \cdot \znorm {\zz ^\prime - \zz}^2, \text{ for all } \zz, \zz ^\prime \in X \times Y \,. 
\) 

In this paper, we refer to a game as a \emph{monotone (general-sum) game} if it satisfies \cref{assumption:convex-concave-combination,assumption:convex average}, and refer to a game as a \emph{strongly monotone} game if it is a monotone game with modulus $\mu, \nu > 0$. It is known that there exists a unique Nash equilibrium for strongly monotone games~\citep{rosen1965existence}. 

\paragraph{Monotone zero-sum games (convex-concave minimax optimization)}
We now consider a subclass: monotone zero-sum games. A two-person game is \emph{zero-sum} if $g=0$. A game is said to be a \emph{monotone zero-sum game} if it is zero-sum and satisfies \cref{assumption:convex-concave-combination}. Note that monotone zero-sum games trivially satisfy \cref{assumption:convex average} (since $g=0$ is convex), and therefore form a subclass of monotone general-sum games. By Sion's minimax theorem~\citep{sion1958minimax}, the NEP for monotone zero-sum games is equivalent to \emph{convex-concave minimax optimization}, that is, finding or approaching a saddle point of the function $h(\cdot, \cdot)$.  
% \[
% \min _{\xx \in X} \ \max _{\yy \in Y} \ h(\xx, \yy) \,. 
% \]

\subsection{Previous results}
\label{sec:previous results}

We begin with a historical overview of the related study of NEPs for monotone games. Many prior studies focus on restrictive cases. Early seminal work by Nesterov and Nemirovski in the early 2000s primarily addressed the restrictive classes of games with equal conditioning ($\mu = \nu$). \cite{nesterov2005smooth} initially studied the bilinearly-coupled, equally-conditioned, and zero-sum cases, which \cite{nemirovski2004prox} later generalized to cover \emph{generally-coupled} and \emph{general-sum} settings. 
Subsequently, research has shifted towards the broader concept of \emph{general conditioning} ($\mu \neq \nu$). This shift is motivated by the need for more flexible modeling of player behaviors in realistic problems and can lead to substantially faster convergence rates, particularly when one of the player has a better conditioning~\citep{chambolle2011first,lin2020near}. 
Early explorations focus on the bilinearly-coupled and zero-sum cases, developing various primal-dual algorithms in different oracle settings~\citep{chambolle2011first,chen2014optimal,kolmogorov2021one,thekumparampil2022lifted}. More recently, the seminal work of \cite{lin2020near} spurred extensive research into \emph{generally-coupled}, \emph{generally-conditioned}, and zero-sum games~(see \cite{yang2020catalyst,wang2020improved,zhang2022lower,kovalev2022first,boct2023accelerated,lan2023novel,lin2025two}, among others). 

Despite these advancements, we are not aware of any study addressing the more general settings of \emph{generally-coupled}, \emph{generally-conditioned}, and \emph{non-zero-sum} games. For this particularly challenging class, the only established results are the long-standing bounds from \cite{tseng1995linear,nemirovski2004prox}. These bounds, however, remain a huge gap from the optimal rates achieved for the zero-sum cases~\citep{lin2020near,kovalev2022first,lan2023novel}. 

Now, we formally outline the state-of-the-art gradient complexity results of monotone general-sum and zero-sum games within the general settings of \emph{general couplings} and \emph{general conditioning}. To simplify the presentation, we assume strong monotonicity for now in the \cref{sec:previous results}, while the results for non-strongly monotone games are indeed similar (as we will discuss later). 
For general-sum games, the NEPs can be solved using variational inequality methods for the operator $\cF$, leading to the following long-standing gradient complexity: 
\begin{proposition}[\cite{tseng1995linear}]
\label{proposition:general-sum games rates}
    For strongly monotone general-sum games, an $\epsilon$-accurate Nash equilibrium can be found with the number of gradient queries bounded by \( \cO \left( \frac L {\min \{\mu, \nu\}} \cdot \log \left( \frac {D^2} \epsilon \right) \right) \,.\)
\end{proposition}

For the zero-sum cases, the gradient complexity can be significantly improved due to recent advances in minimax optimization. 
% Algorithms developed in \cite{lin2020near,kovalev2022first,carmon2022recapp,thekumparampil2022lifted,lan2023novel} achieve a gradient complexity of $\tilde \cO \left( \frac L {\sqrt{\mu \nu}} \cdot \log \left( \frac 1 \epsilon \right) \right)$ for convex-concave minimax optimization. 
% This gradient complexity is minimax optimal, matching the lower complexity bound established in \cite{zhang2022lower}.\footnote{Some prior works (for instance, \cite{zhang2022lower,wang2020improved,thekumparampil2022lifted}) consider different smoothness parameters $(L_{\xx}, L_{\xx\yy}, L_{\yy})$. In this paper, we take $L = L_{\xx} + L_{\xx\yy} + L_{\yy}$ and do not consider the difference in the smoothness parameters.} 

\begin{proposition}[\cite{lin2020near,kovalev2022first,zhang2022lower,lan2023novel}]
\label{proposition:CCMMO rates}
    For strongly monotone zero-sum games, an $\epsilon$-accurate Nash equilibrium can be found with the number of gradient queries bounded by 
    \(
    \cO \left( \frac L {\sqrt{\mu \nu}} \cdot \log \left( \frac {D^2} \epsilon \right) \right) \,. 
    \)
    This rate is minimax optimal, as $\varOmega \left( \frac L {\sqrt{\mu \nu}} \cdot \log \left( \frac {D^2} \epsilon \right) \right)$ gradient queries are required in general. 
\end{proposition}

\subsection{The new problem class}
\label{subsec:new problem class and main theoretical result}

\paragraph{Theoretical motivation}
As shown in \cref{proposition:general-sum games rates,proposition:CCMMO rates}, a huge gap exists in the gradient complexities for solving NEPs for monotone general-sum games versus zero-sum games. This motivates the exploration of an intermediate problem class that partially bridges this gap. 

\paragraph{Monotone near-zero-sum games}
We introduce the class of \emph{monotone $\delta$-near-zero-sum games}, which naturally interpolates between monotone zero-sum ($\delta=0$)\footnote{In $0$-near-zero-sum game, let Player 1 maximize $\aa_1 ^\top \xx + \bb_1 ^\top \yy - h(\xx, \yy)$ and Player 2 maximize $\aa_2 ^\top \xx + \bb_2 ^\top \yy + h(\xx, \yy)$, respectively. 
The Nash equilibrium in the above game is the same as that in the following zero-sum game: Player 1 maximizes $\aa_1 ^\top \xx - \bb_2 ^\top \yy - h(\xx, \yy)$ and Player 2 maximizes $- \aa_1 ^\top \xx + \bb_2 ^\top \yy + h(\xx, \yy)$. 
} and general-sum ($\delta=L$) games. 

\begin{assumption}[Near-zero-sum]
\label{assumption:delta-near-zero-sum}
    Let $\delta \in [0, L]$ such that the function $g (\cdot, \cdot)$ is $\delta$-smooth. 
\end{assumption}

\begin{mdframed}
% \vspace{-0.6em}
\begin{definition}[\textsc{Monotone Near-Zero-Sum Games}]
    If a two-person general-sum game satisfies \cref{assumption:convex-concave-combination,assumption:convex average,assumption:delta-near-zero-sum}, we call it a \emph{monotone $\delta$-near-zero-sum game}. 
\end{definition}
\end{mdframed}

\section{Algorithm and convergence analysis}
\label{sec:algorithm and convergence analysis}

We first focus on the algorithm for strongly monotone near-zero-sum games in \cref{sec:algorithm,sec:convergence analysis}. Then, in \cref{sec:non-strongly monotone}, we present the results for (non-strongly) monotone near-zero-sum games. 

\subsection{Algorithm}
\label{sec:algorithm}

For the zero-sum classes, \cite{lin2020near,kovalev2022first,carmon2022recapp,lan2023novel} obtained the optimal convergence rate by the \algname{Catalyst} method~\citep{lin2018catalyst}. 
However, for the more general classes of non-zero-sum games, the application of \algname{Catalyst} is complicated by the fact that the regularized minimization transforms the problem into a Stackelberg game, whose solution deviates significantly from a Nash equilibrium~(see \cref{sec:smoothing techniques} for more details). Thus, we are not aware of how the similar smoothing techniques can be applied directly to non-zero-sum games. 

This raises the challenge: \emph{can we leverage the off-the-shelf algorithms designed for zero-sum games to solve the non-zero-sum problems of interest?} 
Now, we introduce our novel algorithm, \ICL~(\algname{ICL}), which overcomes the aforementioned challenge and presents a clean black-box framework to {solve near-zero-sum games by using zero-sum algorithms as an oracle}. 

\paragraph{Potential function}
Our algorithm leverages a natural potential function $\Delta \colon X \times Y \rightarrow \R$ defined as: 
\[
\Delta (\zz) = \max _{\tilde \zz = (\tilde \xx, \tilde \yy) \in X \times Y} \ \underbrace {g(\zz) - g(\tilde \zz)} _\text{jointly convex coupling} + \underbrace {h (\xx, \tilde \yy) - h (\tilde \xx, \yy)} _\text{convex-concave zero-sum}, \text{ for all } \zz = (\xx, \yy) \in X \times Y \,. 
\]
This potential function decomposes into a jointly convex coupling part and a convex-concave zero-sum part. % In monotone zero-sum games, the jointly convex part is $0$ and one can have the faster $\tilde \cO \left( \frac {L} {\sqrt{\mu \nu}} \cdot \log \left( \frac 1 \epsilon \right) \right)$ rate. 
We show below in \cref{proposition:delta is the upper bound for NE approximation error,proposition:delta root iff NE} that minimizing this potential function $\Delta (\cdot)$ is sufficient for finding a Nash equilibrium (with detailed proofs in \cref{sec:proof of propositions}): 
\begin{proposition}
\label{proposition:delta is the upper bound for NE approximation error}
For any $\zz = (\xx, \yy) \in X \times Y$, we have $\Delta (\zz) \ge 0$ and 
\[
2 \Delta (\zz) \ge \max _{\tilde \zz = (\tilde \xx, \tilde \yy) \in X \times Y} \ u_1 (\tilde \xx, \yy) - u_1 (\xx, \yy) + u_2 (\xx, \tilde \yy) - u_2 (\xx, \yy) \,. 
\]
\end{proposition}

\begin{proposition}
\label{proposition:delta root iff NE}
    Let $\zz ^* \in X \times Y$. In monotone games, $\zz ^*$ is the Nash equilibrium iff $\Delta (\zz ^*) = 0$. 
\end{proposition}

\paragraph{Algorithm description}
Our \algname{ICL} algorithm solves the strongly monotone near-zero-sum game by {iteratively linearizing the coupling part}, thereby {transforming the non-zero-sum game into a sequence of strongly monotone zero-sum subproblems}. The pseudocode is presented in \cref{algorithm:ICL}. 

Specifically, at every iteration $t$, we linearize the coupling part in the potential function at $\zz_t$: 
\[
\min _{\zz \in X \times Y} \Delta (\zz) \  
{\leadsto} \  
\min _{\zz \in X \times Y} \max _{\tilde \zz \in X \times Y} \lin {\nabla g(\zz_t), \zz - \tilde \zz} + \frac 1 {2 \eta_t} \left( \norm {\zz - \zz_t}^2 - \norm {\tilde \zz - \zz_t}^2 \right) + h (\xx, \tilde \yy) - h (\tilde \xx, \yy) ,
\]
and note that this minimax optimization can be fully decomposed into two separate problems: 
\[
\begin{aligned}
& \min _{\xx \in X} \ \max _{\tilde \yy \in Y} \ \lin {\nabla g(\zz_t), \left( \xx, - \tilde \yy \right)} + h (\xx, \tilde \yy) + \frac 1 {2 \eta_t} \left( \norm {\xx - \xx_t}^2 - \norm {\tilde \yy - \yy_t}^2 \right) = \phi_t (\xx, \tilde \yy) , \text{ and } \\
& \min _{\yy \in Y} \ \max _{\tilde \xx \in X} \ \lin {\nabla g(\zz_t), \left( - \tilde \xx, \yy \right)} - h (\tilde \xx, \yy) + \frac 1 {2 \eta_t} \left( - \norm {\tilde \xx - \xx_t}^2 + \norm {\yy - \yy_t}^2 \right) = - \phi_t (\tilde \xx, \yy) .
\end{aligned}
\]
Moreover, by Sion's minimax theorem~\citep{sion1958minimax}, (after simple substitutions) these two separate problems unify into a \emph{single} saddle point problem of $\min _{\xx \in X}  \max _{\yy \in Y}  \phi_t(\xx, \yy)$, where $\phi_t$ is defined in \cref{eq:define phi t}. The update $\zz _{t+1} = (\xx _{t+1}, \yy_{t+1})$ is then computed by inexactly solving this \emph{unified} saddle point problem, where the inexactness condition is specified in \cref{eq:subproblem accuracy}. Our algorithm, thus, provides a clean and black-box reduction from near-zero-sum games to zero-sum games.

\begin{figure}
\makebox[\linewidth]
{%
\resizebox{\linewidth}{!}{
    \begin{minipage}{1.05\textwidth}
    \begin{algorithm}[H]
        \caption{\ICL\ (\algname{ICL})}
        \label{algorithm:ICL}
        \begin{algorithmic}[1]
        \Require $\zz _0 = (\xx _0, \yy _0) \in X \times Y$. 
        % \Ensure $\zz _T$ is an $\epsilon$-accurate Nash equilibrium
        \For {$t=0,1,\cdots, T-1$}
            \State Let $\phi _t (\xx, \yy) \defEQ$ 
            \begin{equation}
            \label{eq:define phi t} 
                \lin {\nabla _\xx g(\xx _t, \yy _t), \xx} + \frac 1 {2 \eta _t} \norm {\xx - \xx_t} ^2 + h (\xx, \yy) - \lin {\nabla _\yy g(\xx_t, \yy_t), \yy} - \frac 1 {2 \eta _t} \norm {\yy - \yy_t} ^2 \,. 
            \end{equation}
            \State \label{line:ICL subproblem} Find an inexact solution \( \zz _{t+1} = (\xx_{t+1}, \yy_{t+1}) \in X \times Y \) to \( \min _{\xx \in X} \max _{\yy \in Y} \phi_t(\xx, \yy) \) such that 
            \begin{equation}
            \label{eq:subproblem accuracy}
            \lin {\nabla _\xx \phi _t(\zz_{t+1}), \xx_{t+1} - \xx} - \lin {\nabla _\yy \phi _t(\zz_{t+1}), \yy_{t+1} - \yy} \le \epsilon _{t} , \text{ for all } \xx \in X, \ \yy \in Y \,. 
            \end{equation}
        \EndFor
        \end{algorithmic}
    \end{algorithm}
\end{minipage}
}
}
\end{figure}

\subsection{Convergence analysis}
\label{sec:convergence analysis}

Throughout \cref{sec:convergence analysis}, let $\zz ^* = (\xx ^*, \yy ^*)$ be the (unique) Nash equilibrium for the game. We only present the main proof ideas in this section, and the detailed proofs can be found in \cref{sec:proof of convergence}. 
The core of our convergence analysis is to use the properties of the potential function and derive the following descent lemma, based on which we have the outer loop convergence. 
\begin{lemma}[Descent lemma]
\label{lemma:descent lemma}
    In monotone $\delta$-near-zero-sum games, for $\eta _t \le \frac 1 {\delta}$, \cref{algorithm:ICL} ensures 
    \[
    \left( \frac 1 {2 \eta_t} + \frac {\min\{\mu,\nu\}} {2} \right) \norm {\zz_{t+1} - \zz ^*} ^2 \le \frac 1 {2 \eta_t} \norm {\zz_t - \zz ^*} ^2 + \epsilon _t \,. 
    \]        
\end{lemma}

\begin{lemma}[Outer loop convergence]
\label{lemma:outer loop}
    Let $\eta _t = \eta \in (0, \frac 1 \delta]$, for all $t \in [0, T-1] \cap \mathbb Z$. Denote \( \theta = \frac{\min \{ \mu, \nu \}}{\eta ^{-1} + \min \{ \mu, \nu \}} \). 
    Let $\epsilon _t \le { \frac {\theta \epsilon} {4 \eta}}$, for all $t \in [0, T-1] \cap \mathbb Z$. For strongly monotone $\delta$-near-zero-sum games, if the outer loop iterate \(  
    t \ge \frac 1 {\theta} \log \frac {2 D^2} {\epsilon} 
    \), then \cref{algorithm:ICL} converges to an $\epsilon$-accurate Nash equilibrium, that is, $\znorm {\zz _t - \zz ^*} ^2 \le \epsilon$. 
\end{lemma}
%
% \begin{proof}[Proof Sketch]
%     It follows from unrolling the recursion in \cref{lemma:descent lemma} from $T-1, T-2, \cdots$, to $0$. 
% \end{proof}

For the inner loop, any optimal gradient method for zero-sum games can be used: 
\begin{lemma}[Inner loop complexity~\citep{kovalev2022first,carmon2022recapp,thekumparampil2022lifted,lan2023novel}]
\label{lemma:inner loop}
    Under \cref{assumption:convex-concave-combination} with modulus $\mu, \nu > 0$, at each iteration $t \in [0, T-1] \cap \mathbb Z$, for $\eta _t \ge \frac 1 L$, the inexact solution $(\xx_{t+1}, \yy_{t+1})$ in \cref{eq:subproblem accuracy} of \cref{algorithm:ICL} can be found with the number of gradient queries bounded by 
    % \begin{small}
    \(
    \cO \left( \frac L {\sqrt {\left(\eta_t ^{-1} + \mu\right) \left(\eta_t ^{-1} + \nu\right)}} \cdot \log \left( \frac {L D^2 } { \epsilon_t } \right) \right) \,. 
    \) 
    % \end{small}
\end{lemma}

Combining the outer loop convergence result (\cref{lemma:outer loop}) with the inner loop complexity (\cref{lemma:inner loop}), we obtain the main theoretical result of this paper, the overall gradient complexity of \cref{algorithm:ICL}.

\begin{theorem}[Main theoretical result]
\label{thm:total gradient queries}
    Denote $\eta = \min \left\{ \frac 1 \delta, \frac 1 {\min \{\mu, \nu\}} \right\}$ and \( \theta = \frac{\min \{ \mu, \nu \}}{\eta ^{-1} + \min \{ \mu, \nu \}} \). Let $\eta_t = \eta$ and $\epsilon _t = \frac { \theta \epsilon} { 4 \eta }$, for all $t \in [0,T-1] \cap \mathbb Z$. For strongly monotone $\delta$-near-zero-sum games, for \( 
    T \ge \frac 1 {\theta} \log \frac {2 D^2} {\epsilon} 
    \), the outer loop iterates of \cref{algorithm:ICL} converge to an $\epsilon$-accurate Nash equilibrium with the number of gradient queries bounded by  
    % \begin{small}
    \[
    \cO \left( \left(\frac {L} {\sqrt {\mu \nu}} + \frac {L} {\min \{\mu, \nu\}} \cdot \min \left\{ 1, \sqrt{ \frac {\delta} {\mu + \nu} } \right\} \right) \cdot \log \left( \frac { L {D^2} } {\min \{\mu,\nu\} \cdot \epsilon} \right) \log \left( \frac{{D^2}}{\epsilon} \right) \right) \,. 
    \]
    % \end{small}
\end{theorem}

Finally, we highlight the conditions under which \cref{algorithm:ICL} achieves a faster convergence rate compared to variational inequality methods~\citep{tseng1995linear}.

\begin{remark}[Acceleration in strongly monotone near-zero-sum games]
\label{remark:acceleration condition}
    Consider strongly monotone near-zero-sum games where $\min \{ \mu, \nu \} + \delta = o \left( \max \{ \mu, \nu \} \right)$. The gradient complexity of \cref{algorithm:ICL} is 
    \( \tilde \cO \left( \left( \frac {L} {\sqrt {\mu \nu}} + \frac {L} {\min \{\mu, \nu\}} \cdot \sqrt{ \frac {\delta} {\mu + \nu} } \right) \cdot \log ^2 \left( \frac {D^2} \epsilon \right) \right) \),  
    % or equivalently, \( \tilde \cO \left( \left( \frac {L} {\min \{\mu, \nu\}} \cdot \sqrt{ \frac {\min \{ \mu, \nu \} + \delta} {\mu + \nu} } \right) \cdot \log ^2 \left( \frac {D^2} \epsilon \right) \right) 
    % \),
    which (ignoring logarithms) improves upon the $\cO \left( \frac{L}{\min \{\mu,\nu\}} \cdot \log \left( \frac{{D^2}}{\epsilon} \right) \right)$ rate of variational inequality methods as stated in \cref{proposition:general-sum games rates}. We also remark that for the special case of zero-sum games ($\delta=0$), our rate recovers the optimal $\cO \left( \frac {L} {\sqrt{\mu \nu}} \cdot \log \left(\frac {D^2} \epsilon\right) \right)$ rate as stated in \cref{proposition:CCMMO rates} up to a logarithm term. 
\end{remark}

\subsection{Acceleration in non-strongly monotone near-zero-sum games}
\label{sec:non-strongly monotone}

We first state the known results for non-strongly monotone games in literature. 

\begin{proposition}[\cite{nemirovski2004prox}]
    In monotone general-sum games where $\mu=0$ or $\nu=0$, an $\epsilon$-approximate Nash equilibrium can be found within \( \cO \left( \frac {L {D^2} } {\epsilon} \right) \) gradient queries.  
\end{proposition}
Then, we provide our result, which is obtained by a similar reduction as in \cite{lin2020near,wang2020improved,thekumparampil2022lifted}. The proof can be found in \cref{sec:proof of non-strongly monotone}. 
\begin{corollary}
\label{corollary:our informal non-strongly monotone rates}
    In monotone $\delta$-near-zero-sum games with modulus $\mu$ and $\nu$: 
    \begin{enumerate}[label=(\alph*)]
        \item For $\mu > 0$ and $\nu = 0$, an $\epsilon$-approximate Nash equilibrium can be obtained within \( \tilde \cO \left( \left( \frac {L D_Y} {\sqrt { \mu \epsilon }} + \frac {L D_Y^2} {\epsilon} \cdot \min \left\{ 1, \sqrt{\frac {\delta} {\mu}} \right\} \right) \cdot \log^2 \left( \frac {LD^2} \epsilon \right) \right) \) gradient queries; 
        \item For $\mu = 0$ and $\nu > 0$, an $\epsilon$-approximate Nash equilibrium can be obtained within \( \tilde \cO \left( \left( \frac {L D_X} {\sqrt { \nu \epsilon }} + \frac {L D_X^2} {\epsilon} \cdot \min \left\{ 1, \sqrt{\frac {\delta} {\nu}} \right\} \right) \cdot \log^2 \left( \frac {LD^2} \epsilon \right) \right) \) gradient queries; and 
        \item For $\mu = 0$ and $\nu = 0$, an $\epsilon$-approximate Nash equilibrium can be obtained within \( \tilde \cO \left( \left( \frac {L D_X D_Y} {{ \epsilon }} + \frac {L D^2 } {\epsilon} \cdot \min \left\{ 1, \sqrt{\frac {\delta} {\nicefrac{\epsilon}{D_X^2} + \nicefrac{\epsilon}{D_Y^2}}} \right\} \right) \cdot \log^2 \left( \frac {LD^2} \epsilon \right) \right) \) gradient queries. 
    \end{enumerate}
\end{corollary}

\begin{remark}[Acceleration in non-strongly monotone near-zero-sum games]
    For non-strongly monotone general-sum games, our rate of finding an $\epsilon$-approximate Nash equilibrium (ignoring logarithms) is faster than the $\cO \left( \frac{LD^2}{\epsilon} \right)$ rate in literature: (a)~when $\delta = o(\mu)$ and $\nu = 0$; (b)~when $\mu = 0$ and $\delta = o(\nu)$; or (c)~when $\mu=\nu=0$ and $\nicefrac{\epsilon}{D^2} + \delta = o\left( \nicefrac{\epsilon}{D_X^2} + \nicefrac{\epsilon}{D_Y^2} \right)$. We also remark that for the special case of zero-sum games ($\delta=0$), our rates recover the optimal rates of \cite{lin2020near,wang2020improved} up to logarithm terms. 
\end{remark}

\paragraph{Technical novelty} Our technical contributions include a novel coupling linearization technique and the derivation of a descent lemma (\cref{lemma:descent lemma}). These elements combine to form a clean, general, and powerful black-box reduction. This reduction's key advantage is its ability to solve monotone non-zero-sum problems by treating any off-the-shelf zero-sum algorithm as an oracle. Moreover, the inherent black-box nature of our method also allows for deriving complexity results for other oracle settings. For instance, we present some additional results with proximal oracles in \cref{sec:additional results for other oracle and function classes}. 

\section{Application examples}
\label{sec:application examples}

In this section, we present practical examples of monotone near-zero-sum games. We focus on the application of our approach, while the proof details are presented in \cref{sec:proof of examples}. 

\subsection{Our approach for regularized matrix games}
\label{subsec:matrix games with regularization}

\paragraph{Regularized matrix games}
We demonstrate the applicability of \ICL\ to regularized matrix games. Let $X \subseteq \R^n$ and $Y \subseteq \R^m$ be compact convex sets, and let $\mA, \mB \in \R ^{m \times n}$ with $\mnorm {\mA} \le L$, $\mnorm {\mB} \le L$, and $\mnorm {\frac {\mA + \mB} {2}} \le \beta$. Let $\cR \colon X \times Y \to \R$ be an $L$-smooth regularizer that is $\mu$-strongly concave-$\nu$-strongly convex. 
Let \pone\ maximize
\[ 
\fone (\xx, \yy) = \lin {\mA \xx, \yy} + \cR (\xx, \yy) 
\] 
over $\xx \in X$ and \ptwo\ maximize 
\[
\ftwo (\xx, \yy) = \lin {\mB \xx, \yy} - \cR (\xx, \yy) 
\] 
over $\yy \in Y$. Assume that $\beta \le \frac 1 2 \sqrt {\mu \nu}$, then the game is $\min \{ \frac {\mu} 2, \frac {\nu} 2 \}$-strongly monotone, so classic variational inequality methods yield an $\epsilon$-accurate Nash equilibrium within $\cO \left( \frac {L} {\min \{ \mu , \nu \}} \cdot \log \left( \frac {D^2} \epsilon \right) \right)$ gradient queries~\citep{tseng1995linear}.

Now, we show that our \algname{ICL} algorithm can be applied to get a faster rate by leveraging the near-zero-sum structure. 
Since $- \frac 1 2 (u_1 + u_2) (\cdot, \cdot)$ is not jointly convex, violating \cref{assumption:convex average}, we first apply a ``convex reformulation'' technique. 

\paragraph{Convex reformulation technique}
Specifically, we choose the parameters $\beta_1$ and $\beta_2$ based on the relationship between $2\beta$, $\mu$, and $\nu$: 
(i)~If $2 \beta \le \mu$ and $2 \beta \le \nu$, let $\beta_1 = \beta_2 = \beta$; 
(ii)~if $\mu \le 2 \beta \le \nu$, let $\beta_1 = \frac{\mu}{2}$ and $\beta_2 = \frac{2 \beta^2}{\mu}$; and 
(iii)~if $\nu \le 2 \beta \le \mu$, let $\beta_1 = \frac{2 \beta^2}{\nu}$ and $\beta_2 = \frac{\nu}{2}$. 
With these choices, we always have $\beta _1 \le \frac{\mu}{2}$, $\beta_2 \le \frac{\nu}{2}$, and 
\(
\sqrt {\beta _1 \beta _2} = \beta 
\). 

We then reformulate the problem as follows: \pone\ maximizes
\(
\tilde u_1 (\xx, \yy) = u_1 (\xx, \yy) - \beta _2 \norm {\yy} ^2 
\)
over $\xx \in X$, and \ptwo\ maximizes
\(
\tilde u_2 (\xx, \yy) = u_2 (\xx, \yy) - \beta _1 \norm {\xx} ^2  
\)
over $\yy \in Y$. This reformulated NEP has the same Nash equilibrium as the original.  
Let 
\[
\tilde g (\xx, \yy) = - \lin {\left( \frac {\mA + \mB} {2} \right) \xx, \yy} + \left( \frac {\beta _1} {2} \norm{\xx}^2 + \frac {\beta _2} {2} \norm{\yy}^2 \right) 
\]
and 
\[
\tilde h (\xx, \yy) = - \lin {\left( \frac {\mA - \mB} {2} \right) \xx, \yy} - \left( \cR (\xx, \yy) + \frac {\beta _1} {2} \norm{\xx}^2 - \frac {\beta _2} {2} \norm{\yy}^2 \right) \,. 
\]
Then, $\tilde u_1 = - \tilde g - \tilde h$ and $\tilde u_2 = - \tilde g + \tilde h$. Since $\beta \le \frac 1 2 \sqrt {\mu  \nu }$, by Cauchy-Schwartz inequality, $\tilde g(\cdot, \cdot)$ is jointly convex. Further, by the choices of $\beta _1$ and $\beta _2$, we have $\tilde g(\cdot, \cdot)$ is $\left( \beta + \max \{ \beta_1, \beta_2 \} \right)$-smooth, and $\tilde h(\cdot,\cdot)$ is $\frac {\mu} 2$-strongly convex-$\frac {\nu} 2$-strongly concave. 

\paragraph{Our approach applied to reformulated games}
Now applying \cref{algorithm:ICL} to the reformulated NEP, by \cref{thm:total gradient queries}, we obtain an $\epsilon$-accurate Nash equilibrium with the number of gradient queries bounded by 
\[
\tilde \cO \left( \left( \frac {L} {\sqrt {\mu  \nu }} + \frac {L} {\min \{ \mu , \nu  \}} \cdot \frac {\beta} { \sqrt {\mu  \nu } } \right) \cdot \log ^2 \left( \frac {D^2} \epsilon \right) \right) \,. 
\] 
When $\min \{ \mu, \nu \} + \beta = o \left( \frac 1 2 \sqrt {\mu  \nu } \right)$, this rate surpasses the best-known $\cO \left( \frac {L} {\min \{ \mu , \nu  \}} \cdot \log \left( \frac {D^2} \epsilon \right) \right)$ gradient complexity of variational inequality methods~\citep{tseng1995linear}. {This acceleration leveraging the near-zero-sum structure seems to be a new result even in the well-studied context of matrix games.} 

\begin{example}[Matrix games with transaction fees]
\label{example:transactioned matrix games}
    Consider regularized matrix games with transaction fees. Let $X = \cP_n \defEQ \left\{ \xx \in \R _{\ge 0} ^n \mid x_1 + \cdots + x_n = 1 \right\}$ and $Y = \cP_m \defEQ \left\{ \yy \in \R _{\ge 0} ^m \mid y_1 + \cdots + y_m = 1 \right\}$. Let $\mM \in \R ^{m \times n}$ be the payoff matrix of \pone\ \emph{without transaction fee}, with $-\mM$ as the payoff matrix of \ptwo\ \emph{without transaction fee}. Assume $\norm {\mM} \le L$.  
    Denote 
    \( \mM_+ \defEQ \frac 1 2 (\mM + \abs {\mM} ) \text{ and } \mM_- \defEQ \frac 1 2 (-\mM + \abs {\mM} ) 
    \).\footnote{$\abs {\mM}$ represents a matrix of the same dimensions as $\mM$ where each element is the absolute value of the corresponding element in $\mM$. A more concrete illustration is given in \cref{sec:illustration}.} 
    
    Now, suppose there is a \emph{transaction fee} of $\rho \in [0,1]$ charged by some third party on every payment. Then, the payoff matrices of \pone\ and \ptwo\ \emph{with transaction fees} are 
    \[ \mA = (1-\rho) \mM_+ - \mM_- \quad \text{and} \quad \mB = -\mM_+ + (1-\rho) \mM_- \,. \] 
    Let $\cR \colon X \times Y \to \R$ be an $L$-smooth regularizer that is $\mu$-strongly concave-$\nu$-strongly convex. Assume that $\rho \norm {\abs \mM} = o \left( \sqrt{\mu \nu} \right)$.
    
    Let \pone\ maximize 
    \(
    \fone (\xx, \yy) = \lin {\mA \xx, \yy} + \cR (\xx, \yy) 
    \) over $\xx \in X$, 
    and \ptwo\ maximize 
    \(
    \ftwo (\xx, \yy) = \lin {\mB \xx, \yy} - \cR (\xx, \yy)  
    \) over $\yy \in Y$. 
    Applying the convex reformulation and then \cref{algorithm:ICL}, we obtain an $\epsilon$-accurate Nash equilibrium with the number of gradient queries bounded by 
    \[ 
    \tilde \cO \left( \left( \frac {L} {\sqrt {\mu  \nu }} + \frac {L} {\min \{ \mu , \nu  \}} \cdot \frac {\rho \norm {\abs \mM}} { \sqrt {\mu  \nu } } \right) \cdot \log ^2 \left( \frac {D^2} \epsilon \right) \right) \,. 
    \]
    \blue{Matrix games with transaction fees are analogous to many practical scenarios, for instance, (a)~betting sites often set odds such that there is overall no loss-gain~(zero-sum game) but charge a small transaction fee; or (b)~negotiation between two clients via a trusted-third party that charges a brokerage fee is another example of a case with small transaction fee.
    }
\end{example}

\subsection{Our approach for competitive games with small additional incentives}

We show the applicability of \ICL\ to competitive games with small additional incentives. Let $X$ and $Y$ be compact convex sets in Euclidean spaces. Let $h \colon X \times Y \rightarrow \R$ be the competition payoff function, which is $L$-smooth and $\mu$-strongly convex-$\nu$-strongly concave. Let $g \colon X \times Y \rightarrow \R$ be the additional incentive function, which is $\beta$-smooth with $\beta \le L$. Let \pone\ maximize 
\( 
\fone = -g - h 
\) 
over $\xx \in X$, and \ptwo\ maximize 
\(
\ftwo = -g + h 
\) 
over $\yy \in Y$. 
% When the game is $\min \{ \frac {\mu} 2, \frac {\nu} 2 \}$-strongly monotone, classic variational inequality methods yield an $\epsilon$-accurate Nash equilibrium within $\tilde \cO \left( \frac {L} {\min \{ \mu , \nu \}} \cdot \log \left( \frac 1 \epsilon \right) \right)$ gradient queries~\citep{tseng1995linear}. 

% Assume that the general conditioning that $\min \{ \mu, \nu \} = o \left( \max \{ \mu, \nu \} \right)$. 
We explore two scenarios where the games are $\min \{ \frac {\mu} 2, \frac {\nu} 2 \}$-strongly monotone, to which our \algname{ICL} algorithm as well as the classic variational inequalities~\citep{tseng1995linear} can be applied: 
\begin{enumerate}
    \item \label{scenario:jointly convex coupling} If $g(\cdot, \cdot)$ is jointly convex and $\beta = o \left( \max \{ \mu, \nu \} \right)$, applying \cref{algorithm:ICL} directly yields an $\epsilon$-accurate Nash equilibrium with the number of gradient queries bounded by \[ \tilde \cO \left( \left(\frac {L} {\sqrt {\mu \nu}} + \frac {L} {\min \{\mu, \nu\}} \cdot \sqrt{ \frac {\beta} {\mu + \nu} } \right) \cdot \log^2 \left( \frac {D^2} \epsilon \right) \right) \,. \]
    \item \label{scenario:non-convex coupling} If $\beta = o \left( \frac 1 2 \min \{ \mu, \nu \} \right)$, we first apply the ``convex reformulation'' technique. We reformulate the problem as follows: \pone\ maximizes
    \(
    \tilde u_1 (\xx, \yy) = u_1 (\xx, \yy) - \beta \norm {\yy} ^2 
    \)
    over $\xx \in X$, and \ptwo\ maximizes
    \(
    \tilde u_2 (\xx, \yy) = u_2 (\xx, \yy) - \beta \norm {\xx} ^2 
    \)
    over $\yy \in Y$. This reformulated NEP has the same Nash equilibrium as the original. 
    Let 
    \(
    \tilde g = - \frac 1 2 (\tilde u_1 + \tilde u_2) \text{ and } \tilde h = \frac 1 2 (-\tilde u_1 + \tilde u_2) \,. 
    \)
    Then, $\tilde h(\cdot,\cdot)$ is $\frac {\mu} 2$-strongly convex-$\frac {\nu} 2$-strongly concave, and $\tilde g(\cdot, \cdot)$ is jointly convex and $2 \beta$-smooth. 
    
    Applying \cref{algorithm:ICL} to the reformulated NEP, we obtain an $\epsilon$-accurate Nash equilibrium with the number of gradient queries bounded by 
    \[
    \tilde \cO \left( \frac {L} {\sqrt {\mu  \nu }} \cdot \log ^2 \left( \frac {D^2} \epsilon \right) \right) \,. 
    \] 
\end{enumerate}
In both scenarios~\ref{scenario:jointly convex coupling}~and~\ref{scenario:non-convex coupling}, our gradient queries are fewer than the $\cO \left( \frac {L} {\min \{ \mu , \nu \}} \cdot \log \left( \frac {D^2} \epsilon \right) \right)$ gradient queries of the classic variational inequality methods~\citep{tseng1995linear}. 
% We remark that, in scenario~\ref{scenario:non-convex coupling}, the strict conditioning of $\beta = o \left( \frac 1 2 \min \{ \mu, \nu \} \right)$ is required and is in sharp contrast to the weaker conditioning of $\beta = o \left( \frac 1 2 \sqrt { \mu \nu } \right)$ in the ``convex reformulation'' in \cref{subsec:matrix games with regularization}, where the Cauchy-Schwartz inequality can be used under the bilinear coupling structure. \todo{Remove this final remark.}

\begin{example}[Competitive games with small cooperation incentives]
Consider the games where cooperation coexists with competition. Let $X \subseteq X_{\coo} \times X_{\com}$ and $Y \subseteq Y_{\coo} \times Y_{\com}$ be compact convex sets in Euclidean spaces. For $\xx = (\xx _{\coo}, \xx _{\com}) \in X$, $\xx _{\coo} \in X_{\coo}$ represents \pone's effort in cooperation, and $\xx _{\com} \in X _{\com}$ represents \pone's effort in competition (and similarly for $\yy = (\yy _{\coo}, \yy _{\com}) \in Y$). 
Let $f_{\coo} \colon X_{\coo} \times Y_{\coo} \rightarrow \R$ be the cooperation incentive function given by 
\[ 
f_{\coo} (\xx_{\coo}, \yy_{\coo}) = \cR_1 (\xx_{\coo}) + \tilde g(\xx_{\coo}, \yy_{\coo}) + \cR_2 (\yy_{\coo}) \,, 
\] 
where the regularizer $\cR_1\colon X_{\coo} \to \R$ is $\mu$-strongly convex and $L$-smooth, the function $\tilde g\colon X_{\coo} \times Y_{\coo} \to \R$ is jointly convex and $\beta$-smooth, and the regularizer $\cR_2\colon Y_{\coo} \to \R$ is $\nu$-strongly convex and $L$-smooth. Let $f_{\com} \colon X_{\com} \times Y_{\com} \rightarrow \R$ be the competition payoff function, which is $L$-smooth and $\mu$-strongly convex-$\nu$-strongly concave. Assume that $\beta = o \left( \max \{ \mu , \nu \} \right)$. 

Let \pone\ maximize 
\( 
\fone (\xx, \yy) = - f_{\coo} (\xx_{\coo}, \yy_{\coo}) - f_{\com} (\xx_{\com}, \yy_{\com})
\) 
over $\xx \in X$, and \ptwo\ maximize 
\(
\ftwo (\xx, \yy) = - f_{\coo} (\xx_{\coo}, \yy_{\coo}) + f_{\com} (\xx_{\com}, \yy_{\com})
\) 
over $\yy \in Y$. Denoting $\tilde h(\xx, \yy) = \cR_1 (\xx_{\coo}) + f_{\com} (\xx_{\com}, \yy_{\com}) - \cR_2 (\yy_{\coo})$, the NEP can be reformulated as \pone\ maximizing 
\(
\tilde u_1 (\xx, \yy) = - \tilde g (\xx_{\coo}, \yy_{\coo}) - \tilde h (\xx, \yy)
\)
over $\xx \in X$, and \ptwo\ maximizing 
\(
\tilde u_2 (\xx, \yy) = - \tilde g (\xx_{\coo}, \yy_{\coo}) + \tilde h (\xx, \yy)
\)
over $\yy \in Y$. Applying \cref{algorithm:ICL} as detailed above, we obtain an $\epsilon$-accurate Nash equilibrium with the number of gradient queries bounded by 
\[ 
\tilde \cO \left( \left( \frac {L} {\sqrt {\mu \nu}} + \frac {L} {\min \{\mu, \nu\}} \cdot \sqrt{ \frac {\beta} {\mu + \nu} } \right) \cdot \log^2 \left( \frac {D^2} \epsilon \right) \right) \,. 
\]
As a final remark, the modeling of the coexistence of competition and cooperation has been well-researched (for instance, see the studies of \cite{nash1950bargaining,nash1953two,selten1960bewertung,raiffa1952arbitration,kalai1978arbitration,kalai2013cooperation,halpern2013cooperative} on semi-cooperative games). Indeed, these theories are often applied to the scenarios where cooperation dominates, and optimization techniques have been used to accelerate the dominant cooperation part~\citep{chen2017accelerated}. Our work contributes to this line of research on semi-cooperative games where competition dominates, yet there is a small cooperation incentive. 
\blue{
For example, repeated Prisoners' dilemma with a stochastic number of rounds with small benefit-to-cost ratio is an example of competition with small cooperation incentive~(as benefit-to-cost ratio $b/c$ is small)~\citep{nowak2006evolutionary,nowak2006five,sigmund2010calculus}. We give a more concrete illustration of donation games in \cref{sec:illustration}.
}
%Further exploration of our scenarios in practical applications is an interesting direction for future research. 
\end{example}

\section{Basic  numerical experiments}

We conducted basic numerical experiments to validate our theoretical results, focusing on matrix games with transaction fees as in \cref{example:transactioned matrix games}. We set $n=m=10000$, $\mu=10^{-4}$, $\nu=1$, and $\epsilon=10^{-7}$. A sparse, random matrix $\mM \in \R ^{m \times n}$ such that $\norm {\mM} = 1$ was generated. The regularizer was defined as $\cR (\xx, \yy) = - \frac {\mu} 2 \norm {\xx } ^2 + \frac {\nu} 2 \norm {\yy } ^2$. We varied the transaction fee $\rho$ from $\{ 0.00\%, 0.03\%, \cdots, 0.18\% \}$. 
Our implementation of \algname{ICL}~(\cref{algorithm:ICL}), detailed in \cref{subsec:matrix games with regularization}, used the Lifted Primal-Dual method~\citep{thekumparampil2022lifted} for the inner loop. We compared \algname{ICL} against the Optimistic Gradient Descent Ascent~(\algname{OGDA})~\citep{popov1980modification} and Extra-Gradient~(\algname{EG})~\citep{korpelevich1976extragradient} methods for variational inequalities. More details and additional experiments are provided in \cref{sec:more details and additional experiments}, and our code is available at \url{https://github.com/riekenluo/Monotone_Near_Zero_Sum_Games}. 

\begin{table}[ht]
\centering
\caption{Gradient query counts (in thousands) to converge to an $\epsilon$-accurate Nash equilibrium under various transaction fees. Error bars indicate $2$-sigma variations across $10$ independent runs.}
\label{tab:results under different transaction rates}
\resizebox{\linewidth}{!}{%
    \begin{tabular}{cccccccc}
      \toprule
      \diagbox[width=12em]{Methods}{Transaction Fee $\rho$} & $0.00\%$ & $0.03\%$ & $0.06\%$ & $0.09\%$ & $0.12\%$ & $0.15\%$ & $0.18\%$ \\
      \midrule
      \algname{ICL}~(\cref{algorithm:ICL}) & $ \mathbf{9.1 \pm 0.0} $ & $ \mathbf{22.6 \pm 0.4} $ & $ \mathbf{42.2 \pm 0.3} $ & $ \mathbf{65.0 \pm 0.3} $ & $ \mathbf{75.7 \pm 0.3} $ & $ 113.7 \pm 0.7 $ & $ 123.8 \pm 0.6 $ \\
      \algname{OGDA}~\citep{popov1980modification} & $ 93.9 \pm 0.5 $ & $ 93.9 \pm 0.5 $ & $ 93.9 \pm 0.5 $ & $ 93.9 \pm 0.5 $ & $ 93.9 \pm 0.5 $ & $ \mathbf{94.0 \pm 0.6} $ & $ \mathbf{94.0 \pm 0.6} $ \\
      \algname{EG}~\citep{korpelevich1976extragradient} & $ 132.9 \pm 0.8 $ & $ 132.9 \pm 0.8 $ & $ 132.9 \pm 0.8 $ & $ 132.9 \pm 0.8 $ & $ 132.9 \pm 0.8 $ & $ 132.9 \pm 0.8 $ & $ 132.9 \pm 0.8 $ \\
      \bottomrule
    \end{tabular}
}
\end{table}

\blue{The numerical results, summarized in \cref{tab:results under different transaction rates}, demonstrate that: (i)~our \algname{ICL} algorithm converges faster when the game is closer to a zero-sum one; (ii)~the classic variational inequality methods (\algname{EG} or \algname{OGDA}) do not benefit from the near-zero-sum structure; and (iii)~our \algname{ICL} algorithm is faster than the classic variational inequality methods when the game is sufficiently near-zero-sum.} 
In particular, in the experiments, \algname{ICL} requires fewer gradient queries to converge to an $\epsilon$-accurate Nash equilibrium when the transaction fee $\rho \le 0.12\%$. This empirical observation aligns with our theoretical prediction in \cref{example:transactioned matrix games}, which suggests that \algname{ICL} converges faster when $\rho \norm {\abs {\mM}} \lll \sqrt {\mu \nu} = 1 \%$. 

\section{Conclusions, limitations, and future work}
\label{sec:conclusions}
In this work we consider the class of monotone games and present a condition that naturally interpolates between the zero-sum and a non-zero-sum class. We develop an efficient gradient-based approach and show its applicability with several examples motivated from the literature. 

There are some limitations of our work:
(a)~in our complexity there is a $\log^2 (\frac {D^2} \epsilon)$ dependency rather than a single logarithm dependency, and whether this double logarithm dependency can be removed is an interesting question; and
(b)~whether lower-bound results can be obtained for the new class also remains a challenging question \blue{ and may involve difficult construction of non-quadratic functions, especially given that the lower bound for the general-sum classes remains widely open relative to the long-standing upper bound established in \cite{tseng1995linear,nemirovski2004prox}. 

The current work focuses on two-player monotone non-zero-sum games, which constitutes a crucial first step in generalizing two-player monotone zero-sum games with general conditioning. 
A natural, though highly non-trivial, extension involves generalizing our fast convergence rates to the multiplayer near-zero-sum setting. 
We notice that, as a preliminary step, the fast rate in zero-sum games~(\cref{proposition:CCMMO rates}) have not yet been extended to multiplayer settings. 
Indeed, this preliminary step is conceptually challenging, as the zero-sum condition in multiplayer games does not imply the strict competition found in two-player games. 
Furthermore, classic reductions in \cite{von1947theory} show that an $n$-player general-sum game can be reduced to an $(n+1)$-player zero-sum game by letting the $(n+1)$th player take the negative of the summation of the first $n$ players. This implies that three-player zero-sum games are inherently no easier than two-player general-sum games. 
These results suggest that achieving a fast convergence rate in the multiplayer setting will likely require significantly stronger structural assumptions than the simple summation of utilities to zero. 
Hence, we leave the full generalization to the multiplayer setting for future research.}

In addition to the above theoretical aspects, there are several other interesting directions as well: for example,
(a)~exploring other applications of regularized matrix games with near-zero-sum payoff matrices is an interesting direction; and 
(b)~in the research of semi-cooperative games where competition dominates, applying our methods in more practical examples is another fruitful direction for future research. 

%%%%%%%%%%%%%%%%%%%%%%%%%%%%%%%%%%%%%%%%%%%%%%%%%%%%%%%%%%%%

\subsubsection*{Acknowledgments}
The authors thank for the helpful discussion with Anton Rodomanov during the initial preparation of this project. 

RL and KC acknowledge the support of ERC CoG 863818 (ForM-SMArt) and Austrian Science Fund (FWF) 10.55776/COE12. 

% \clearpage

% \paragraph{Ethics statement}

% This is a theoretical research without any ethics concern. All the experiments are based on randomly generated data. The only LLM usage is to aid or polish writing. 

% \paragraph{Reproducibility statement}

% We include the detailed proofs of all the results in the Appendix. We include the source codes of all the experiments, including the scripts for the generation of random data. The codes are based on standard Python libraries (like \texttt{numpy} or \texttt{scipy}) and are easy to run. 

\bibliography{reference}
\bibliographystyle{iclr2026_conference}

\clearpage
\appendix

\appendix

\section{Related work}
\label{sec:related work}

We discuss other classes of games in the literature that bridge the gap between zero-sum and general-sum games. 

\paragraph{Near network zero-sum games}
Near network zero-sum games~\citep{hussain2023beyond} define a class of games that is close to network zero-sum games in terms of maximum pairwise difference~\citep{candogan2013dynamics,hussain2023beyond}. Limited to the setting of two-person games, monotone near-zero-sum games considered in this paper differ in three aspects: (i)~The utility functions in this paper can be general functions, rather than bilinear functions; 
(ii)~the difference between near-zero-sum games and zero-sum games in this paper is characterized by (higher-order) smoothness parameter, rather than by function values; and (iii) the solution of near network zero-sum games is taken directly from the zero-sum case, which only guarantees convergence to a neighborhood of the Nash equilibrium. 

% \paragraph{Near-potential games}
% Potential games~\citep{monderer1996potential} model strict cooperation in multi-player settings, where the utility functions of the players are multi-linear. As a generalization of potential games, \cite{candogan2013dynamics} introduce $\tau$-near-potential games, which are ``close'' to potential games in the sense that their maximum pairwise difference is bounded by $\tau$. Limited to the setting of two-person games, monotone near-zero-sum games considered in this paper differ from near-potential games in three aspects: 
% (i) The utility functions in this paper can be general functions, rather than bilinear functions; 
% (ii) The difference between near-zero-sum games and zero-sum games is characterized by (higher-order) smoothness parameter, rather than by function values; and 
% (iii) Most importantly, (near-)potential games primarily model cooperation-dominated scenarios, a focus that stands in direct contrast to the competition-dominated scenarios considered in this paper. 

\paragraph{Rank-$k$ games}
In the setting of matrix games, one of the most significant attempts on bridging the gap between zero-sum and non-zero-sum games is the class of Rank-$k$ games introduced in \cite{kannan2010games}. As a generalization of zero-sum matrix games, \cite{kannan2010games} study matrix games where $\operatorname{rank}(\mA + \mB) = k$, where $\mA$ and $\mB$ are the payoff matrices of the two players. To find an approximate Nash equilibrium, an FPTAS exists when $k$ is small~\citep{kannan2010games}; to find an exact Nash equilibrium, Rank-$1$ games can be solved in polynomial time~\citep{adsul2021fast}, while Rank-$3$ games are already PPAD-hard~\citep{mehta2018constant}. 
It is crucial to emphasize that monotone near-zero-sum games, as considered in this paper, are fundamentally distinct from Rank-$k$ games. Specifically: 
(i)~The utility functions in this paper can be general functions, rather than bilinear functions; 
(ii)~matrix games can be sufficiently near-zero-sum but still have full rank; and
(iii)~the focus of this paper is on gradient-based algorithms and complexity within the Nemirovsky-Yudin optimization model~\citep{nemirovski1983problem}, while the study of Rank-$k$ games focuses on algorithms and complexity on Turing machines.

\section{Relations to approximate Nash equilibrium}
\label{sec:different Nash equilibrium}

Indeed, an approximate Nash equilibrium can be obtained from an accurate Nash equilibrium~\citep{nemirovski2004prox}. Below, we include this result for self-consistency. 
\begin{proposition}[\cite{nemirovski2004prox}]
\label{proposition:approximate-gradient-distance}
    In a monotone general-sum game, let $\zz^* = (\xx^*, \yy^*)$ be the Nash equilibrium. Let $\bar \zz = (\bar \xx, \bar \yy) \in X \times Y$ and $\gamma \in (0, \frac 1 {\sqrt 2 L}]$. We have 
    \[ \begin{aligned}
    &\quad \max _{\xx \in X,\ \yy \in Y} \ u_1 (\xx, \hat \yy) - u_1 (\hat \xx, \hat \yy) + u_2 (\hat \xx, \yy) - u_2 (\hat \xx, \hat \yy) \\
    &\le \max _{\xx \in X,\ \yy \in Y} \ \lin {\cF (\hat \xx, \hat \yy), (\hat \xx, \hat \yy) - (\xx, \yy)} \\
    &\le \frac 2 \gamma \sqrt {D_X^2 + D_Y^2} \norm {\bar \zz - \zz^*} \,, 
    \end{aligned} \] 
    where $\hat \xx \defEQ \oldPi _{X} \left( \bar \xx + \gamma \nabla _\xx u_1 (\bar \xx, \bar \yy) \right)$ and $\hat \yy \defEQ \oldPi _{Y} \left( \bar \yy + \gamma \nabla _\yy u_2 (\bar \xx, \bar \yy) \right)$.\footnote{In a Euclidean space $\cQ$, for a non-empty, closed, and convex set $Q \subseteq \cQ$ and a vertex $\uu \in \cQ$, let $\oldPi _{Q} (\uu)$ denote the projection of $\uu$ onto $Q$, that is, $\oldPi _{Q} (\uu) \defEQ \argmin _{\vv \in Q} \norm{\uu - \vv}.$}
\end{proposition}
\begin{proof}
    Denote $\xx _+ \defEQ \oldPi _{X} \left( \bar \xx + \gamma \nabla _{\xx} u_1 (\hat \xx, \hat \yy) \right)$, $\yy _+ \defEQ \oldPi _{Y} \left( \bar \yy + \gamma \nabla _{\yy} u_2 (\hat \xx, \hat \yy) \right)$, and $\zz _+ \defEQ (\xx _+, \yy _+)$. Consider any $\tilde \xx \in X$ and $\tilde \yy \in Y$. 
    By the assignment of $\hat \xx$, we have 
    \begin{equation}
    \label{eq:hat x descent}
    \lin {\nabla _\xx u_1 (\bar \xx, \bar \yy), \hat \xx} - \frac 1 {2 \gamma} \norm {\hat \xx - \bar \xx} ^2 \ge \lin {\nabla _\xx u_1 (\bar \xx, \bar \yy), \xx_+} - \frac 1 {2 \gamma} \norm {\xx_+ - \bar \xx} ^2 + \frac 1 {2 \gamma} \norm {\xx_+ - \hat \xx} ^2 \,. 
    \end{equation}
    By the assignment of $\xx _+$, we have
    \begin{equation}
    \label{eq:x + descent}
    \lin {\nabla _\xx u_1 (\hat \xx, \hat \yy), \xx _+ } - \frac 1 {2 \gamma} \norm {\xx _+ - \bar \xx} ^2 \ge \lin {\nabla _\xx u_1 (\hat \xx, \hat \yy), \tilde \xx } - \frac 1 {2 \gamma} \norm {\bar \xx - \tilde \xx} ^2 + \frac 1 {2 \gamma} \norm {\xx _+ - \tilde \xx} ^2 \,. 
    \end{equation}

    In view of
    \begin{equation}
    \label{eq:hat x upper bound}
    \begin{aligned}
        &\quad \lin { \nabla _\xx u_1(\hat \xx, \hat \yy), \tilde \xx - \hat \xx } \\
        &= \lin { \nabla _\xx u_1 (\hat \xx, \hat \yy), \xx _+ - \hat \xx } + \lin { \nabla _\xx u_1 (\hat \xx, \hat \yy), \tilde \xx - \xx _+} \\
        &\refLE{eq:x + descent} \lin { \nabla _\xx u_1 (\hat \xx, \hat \yy), \xx _+ - \hat \xx } - \frac 1 {2 \gamma} \norm {\xx _+ - \bar \xx}^2 + \frac 1 {2 \gamma} \norm {\bar \xx - \tilde \xx}^2 - \frac 1 {2 \gamma} \norm {\xx _+ - \tilde \xx}^2 \\
        &= \lin { \nabla _\xx u_1 (\hat \xx, \hat \yy) - \nabla _\xx u_1 (\bar \xx, \bar \yy), \xx _+ - \hat \xx } + \lin { \nabla _\xx u_1 (\bar \xx, \bar \yy), \xx _+ - \hat \xx } - \frac 1 {2 \gamma} \norm {\xx _+ - \bar \xx}^2 \\
        &\qquad + \frac 1 {2 \gamma} \norm {\bar \xx - \tilde \xx}^2 - \frac 1 {2 \gamma} \norm {\xx _+ - \tilde \xx}^2 \\
        &\refLE{eq:hat x descent} \lin { \nabla _\xx u_1 (\hat \xx, \hat \yy) - \nabla _\xx u_1 (\bar \xx, \bar \yy), \xx _+ - \hat \xx } - \frac 1 {2 \gamma} \norm {\hat \xx - \bar \xx} ^2 - \frac 1 {2 \gamma} \norm {\xx_+ - \hat \xx} ^2 \\
        &\qquad + \frac 1 {2 \gamma} \norm {\bar \xx - \tilde \xx}^2 - \frac 1 {2 \gamma} \norm {\xx _+ - \tilde \xx}^2 \\ 
        &\le L \norm { (\hat \xx, \hat \yy) - \bar \zz } \cdot \norm {\xx _+ - \hat \xx}  - \frac 1 {2 \gamma} \norm {\hat \xx - \bar \xx} ^2 - \frac 1 {2 \gamma} \norm {\xx_+ - \hat \xx} ^2 \\
        &\qquad + \frac 1 {2 \gamma} \norm {\bar \xx - \tilde \xx}^2 - \frac 1 {2 \gamma} \norm {\xx _+ - \tilde \xx}^2 \\
        &\le \frac{L}{2 \sqrt 2} \norm { (\hat \xx, \hat \yy) - \bar \zz } ^2 - \frac 1 {2 \gamma} \norm {\hat \xx - \bar \xx} ^2 + \frac 1 {2 \gamma} \norm {\bar \xx - \tilde \xx}^2 - \frac 1 {2 \gamma} \norm {\xx _+ - \tilde \xx}^2 
    \end{aligned}
    \end{equation}
    (where we have used $\gamma \le \frac 1 {\sqrt 2 L}$ in the last inequality), and similarly, 
    \begin{equation}
    \label{eq:hat y upper bound}
        \lin { \nabla _\yy u_2(\hat \xx, \hat \yy), \tilde \yy - \hat \yy } 
        \le \frac{L}{2 \sqrt 2} \norm { (\hat \xx, \hat \yy) - \bar \zz } ^2  - \frac 1 {2 \gamma} \norm {\hat \yy - \bar \yy} ^2 + \frac 1 {2 \gamma} \norm {\bar \yy - \tilde \yy}^2 - \frac 1 {2 \gamma} \norm {\yy _+ - \tilde \yy}^2 \,, 
    \end{equation}
    we have 
    \begin{equation}
    \label{eq:hat x hat y upper bound}
    \begin{aligned}
        &\quad \lin {\cF (\hat \xx, \hat \yy), (\hat \xx, \hat \yy) - (\tilde \xx, \tilde \yy)} \\
        &\stackrel{\eqref{eq:hat x upper bound}\eqref{eq:hat y upper bound}}{=} (\frac L {\sqrt 2} - \frac 1 {2 \gamma}) \norm { (\hat \xx, \hat \yy) - \bar \zz } ^2 + \frac 1 {2 \gamma} \norm {\bar \zz - (\tilde \xx, \tilde \yy)}^2 - \frac 1 {2 \gamma} \norm {\zz _+ - (\tilde \xx, \tilde \yy)}^2 \\
        &\le \frac 1 {2 \gamma} \norm {\bar \zz - (\tilde \xx, \tilde \yy)}^2 - \frac 1 {2 \gamma} \norm {\zz _+ - (\tilde \xx, \tilde \yy)}^2 \,, 
    \end{aligned}
    \end{equation}
    where we have used $\gamma \le \frac 1 {\sqrt 2 L}$ in the last inequality. 

    Taking $(\tilde \xx, \tilde \yy) := (\xx^*, \yy^*)$ in \cref{eq:hat x hat y upper bound} for the moment, we get 
    \begin{equation}
    \label{eq:extragradient monotonicity}
        \norm {\zz _+ - \zz^*}^2 \refLE{eq:hat x hat y upper bound} \norm {\bar \zz - \zz^*}^2 - 2 \gamma \lin {\cF (\hat \xx, \hat \yy), (\hat \xx, \hat \yy) - \zz^*} \le \norm {\bar \zz - \zz^*}^2 \,. 
    \end{equation}

    Finally, in view of
    \begin{equation}
    \label{eq:hat x hat y distance bound}
    \begin{aligned}
        &\quad \lin {\cF (\hat \xx, \hat \yy), (\hat \xx, \hat \yy) - (\tilde \xx, \tilde \yy)} \\
        &\refLE{eq:hat x hat y upper bound} \frac 1 {2 \gamma} \norm {\bar \zz - (\tilde \xx, \tilde \yy)}^2 - \frac 1 {2 \gamma} \norm {\zz _+ - (\tilde \xx, \tilde \yy)}^2 \\
        &= \frac 1 {2 \gamma} \left( \norm {\bar \xx - \tilde \xx}^2 - \norm {\xx _+ - \tilde \xx}^2 + \norm {\bar \yy - \tilde \yy}^2 - \norm {\yy _+ - \tilde \yy}^2 \right) \\
        &\le \frac 1 {2 \gamma} \left( \norm {\bar \xx - \tilde \xx + \xx _+ - \tilde \xx} \cdot \norm {\bar \xx - \xx _+} + \norm {\bar \yy - \tilde \yy + \yy _+ - \tilde \yy} \cdot \norm {\bar \yy - \yy _+} \right) \\
        &\le \frac 1 {\gamma} \left( D_X \norm {\bar \xx - \xx _+} + D_Y \norm {\bar \yy - \yy _+} \right) \\
        &\le \frac 1 {\gamma} \sqrt {D_X^2 + D_Y^2} \cdot \sqrt {\norm{\bar \xx - \xx _+}^2 + \norm{\bar \yy - \yy _+}^2} \\
        &\le \frac 1 {\gamma} \sqrt {D_X^2 + D_Y^2} \cdot \sqrt {2 \norm{\bar \zz - \zz ^*}^2 + 2 \norm{\zz _+ - \zz ^*}^2} \\
        &\refLE{eq:extragradient monotonicity} \frac 2 {\gamma} \sqrt {D_X^2 + D_Y^2} \cdot \norm{\bar \zz - \zz ^*} \,,
    \end{aligned}
    \end{equation}
    we have 
    \[
    \begin{aligned}
        u_1 (\tilde \xx, \hat \xx) - u_1 (\hat \xx, \hat \yy) + u_2 (\hat \xx, \tilde \yy) - u_2 (\hat \xx, \hat \yy) 
        &\le \lin {\cF (\hat \xx, \hat \yy), (\hat \xx, \hat \yy) - (\tilde \xx, \tilde \yy)} \\
        &\refLE{eq:hat x hat y distance bound} \frac 2 {\gamma} \sqrt {D_X^2 + D_Y^2} \cdot \norm{\bar \zz - \zz ^*} \,,
    \end{aligned} 
    \]
    and the desired bound follows because $\tilde \xx$ and $\tilde \yy$ can take arbitrary points in $X$ and $Y$, respectively. 
\end{proof}

% A pair of decisions $\bar \zz = (\bar \xx, \bar \yy) \in X \times Y$ is called $\epsilon$-\emph{stationary point} with stepsize $\gamma$, if 
% \[
% \norm {\bar \zz - \oldPi _Z \left( \bar \zz - \gamma \cF (\bar \zz) \right)} \le \gamma \epsilon \,.
% \]

We also state the following sufficient condition for the accurate Nash equilibrium, which can be used as stopping criterion for the optimization algorithms. Similar results can be found, for instance, in \cite{nemirovski2004prox,yang2020catalyst}. 
\begin{proposition}[Stopping criterion~\cite{nemirovski2004prox,yang2020catalyst}]
\label{proposition:extra gradient stopping criterion}
    In a monotone general-sum game, let $\zz^* = (\xx^*, \yy^*)$ be the Nash equilibrium. Let $\bar \zz = (\bar \xx, \bar \yy) \in X \times Y$, $\gamma \in (0, \frac 1 {2 L}]$, and $\mu = \min \{ \mu, \nu \}$. We have 
    \[ 
    \norm {\bar \zz - \zz^*}^2 \le \left( \frac 4 {\mu^2 \gamma^2} - \frac{2}{\mu \gamma} + 16 \right) \norm {\zz_+ - \bar \zz} ^2 \,,
    \] 
    where $\zz _+ = \oldPi _Z \left( \bar \zz - \gamma \cF (\hat \zz) \right)$, in which $\hat \zz = \oldPi _Z \left( \bar \zz - \gamma \cF (\bar \zz) \right)$. 
\end{proposition}
\begin{proof} 
    We have
    \[
    \begin{aligned}
        &\quad (1- \frac{\mu \gamma}{2}) \norm {\bar \zz - \zz^*}^2 - \left( \frac{2}{\mu \gamma} - 1 \right) \norm {\zz _+ - \bar \zz} ^2 \\
        &\le \norm { \zz _{+} - \zz^* }^2 \\
        &\refLE{eq:hat x hat y upper bound} \norm { \bar \zz - \zz^* } ^2 - 2 \gamma \lin {\cF (\hat \zz), \hat \zz - \zz^*} \\
        &\le \norm { \bar \zz - \zz^* } ^2 - 2 \gamma \lin {\cF (\hat \zz) - \cF (\zz^*), \hat \zz - \zz^*} \\
        &\le \norm { \bar \zz - \zz^* } ^2 - 2 \mu \gamma \norm {\hat \zz - \zz ^*} ^2 \\
        &\le \norm { \bar \zz - \zz^* } ^2 - \mu \gamma \norm {\bar \zz - \zz ^*} ^2 + 2 \mu \gamma \norm {\hat \zz - \bar \zz} ^2 \\
        &= (1 - \mu \gamma) \norm {\bar \zz - \zz^*} ^2 - 2 \mu \gamma \norm {\hat \zz - \bar \zz} ^2 + 4 \mu \gamma \norm {\hat \zz - \bar \zz} ^2 \\
        &\le (1 - \mu \gamma) \norm {\bar \zz - \zz^*} ^2 - 2 \mu \gamma \norm {\hat \zz - \bar \zz} ^2 + 8 \mu \gamma \norm {\zz_+ - \bar \zz}^2 + 8 \mu \gamma \norm {\zz_+ - \hat \zz}^2 \\
        &\le (1 - \mu \gamma) \norm {\bar \zz - \zz^*} ^2 - 2 \mu \gamma \norm {\hat \zz - \bar \zz} ^2 + 8 \mu \gamma \norm {\zz_+ - \bar \zz}^2 + 8 \mu \gamma \norm {\bar \zz - \gamma \cF (\hat \zz) - \bar \zz - \gamma \cF (\bar \zz)}^2 \\
        &\le (1 - \mu \gamma) \norm {\bar \zz - \zz^*} ^2 - 2 \mu \gamma \norm {\hat \zz - \bar \zz} ^2 + 8 \mu \gamma \norm {\zz_+ - \bar \zz}^2 + 8 \mu L^2 \gamma^3 \norm {\hat \zz - \bar \zz}^2 \\
        &\le (1 - \mu \gamma) \norm {\bar \zz - \zz^*} ^2 + 8 \mu \gamma \norm {\zz_+ - \bar \zz}^2 \,, 
    \end{aligned}
    \]
    where in the second to last inequality we use $\gamma \le \frac 1 {2L}$. Finally, the desired bound follows from rearrangement. 
\end{proof}

\section{Discussions on the Catalyst methods}
\label{sec:smoothing techniques}

In this section, we present the intuition of most existing algorithms for convex-concave minimax optimization considering general conditioning, and explain why similar idea may not work directly when generalized to monotone near-zero-sum games. 

Most of existing algorithms for minimax optimization with general conditioning are based on \algname{Catalyst}~\citep{lin2018catalyst}. In minimax optimization, we have $\fone + \ftwo = 0$. Assume without loss of generality that $\mu \le \nu$. The function $f(\xx) \defEQ - \fone (\xx, \yy(\xx))$ is $\mu$-strongly convex over $\xx \in X$, in which $\yy (\xx) \defEQ \argmax _{\yy \in Y} \ftwo (\xx, \yy)$. At the core of these algorithms, they build a function $\hat {f} _{t}$ and get an inexact solution $\hat \xx _{t+1}$ at each iteration $t$: 
\begin{equation}
\label{eq:smoothing}
\hat \xx _{t+1} \approx \argmin _{\xx \in X}\ \left[ \hat {f} _{t} (\xx) \defEQ f (\xx) + \frac {\nu} {2} \xnorm {\xx - \hat \xx _{t}} ^{2} \right] \,. 
\end{equation}
The outer loop is an inexact accelerated proximal point algorithm with $\tilde \cO \left( \sqrt {\frac {\nu} {\mu}} \cdot \log \left( \frac 1 \epsilon \right) \right)$ iterations~\citep{nesterov2005smooth,lin2018catalyst,carmon2022recapp}, and the inner loop of solving the smoothed \cref{eq:smoothing} can be any method with the number of gradient queries $\tilde \cO \left( \frac {L} {\nu} \cdot \log \left( \frac 1 \epsilon \right) \right)$~\citep{tseng1995linear}. So, the total gradient complexity is\footnote{The double logarithm term may be avoided by combining this algorithmic idea with some complicated techniques~\citep{kovalev2022first,carmon2022recapp}, which we omit here for the simplicity of presentation.}
\[ 
\tilde \cO \left( \sqrt {\frac {\nu} {\mu}} \cdot \log \left( \frac 1 \epsilon \right) \right) \cdot \tilde \cO \left( \frac {L} {\nu} \cdot \log \left( \frac 1 \epsilon \right) \right) =  \tilde \cO \left( \frac L {\sqrt {\mu \nu}} \cdot \log ^2 \left( \frac 1 \epsilon \right) \right) \,. 
\] 

However, if we try to apply the above \algname{Catalyst} method to monotone non-zero-sum games, the algorithm may only converge to a Stackelberg solution, which can be very different from the Nash equilibrium in non-zero-sum games. 

\begin{example}[Stackelberg solution] 
Consider the case where $X = [0, 1] \times [1, 2] \subseteq \R^2$ and $Y = [-1, 0] \subseteq \R$. Let \pone\ maximize 
\[ u_1 (\xx, y) = - \frac 1 2 (x_1 - 1) ^2 - \frac 1 2 (x_2 - 1) ^2 + \frac 1 2 x_1 y \] over $\xx \in X$, and \ptwo\ maximize 
\[ 
u_2 (\xx,y) = \frac 1 2 x_2 y - (y+1) ^2 
\] over $y \in Y$. 
Then, the \algname{Catalyst} minimization of $f(x) = - u_1 (\xx, y(\xx))$ will lead to the Stackelberg solution $(\xx= \left( \frac {40} {63}, \frac {68} {63} \right), \ y = - \frac{46}{63})$, which is different from the Nash equilibrium $(\xx= \left( \frac 5 8, 1 \right),\ y=-\frac 3 4)$. 
\end{example}
Therefore, we are not aware of how the \algname{Catalyst} methods for convex-concave minimax optimization can be applied in NEPs for non-zero-sum games. 

\section{Proof details}
\label{sec:proof details}

\subsection{Proofs for the results in \texorpdfstring{\Cref{sec:algorithm}}{}}
\label{sec:proof of propositions}

\begin{proof}[Proof of \cref{proposition:delta is the upper bound for NE approximation error}]
    For any $\zz = (\xx, \yy) \in X \times Y$, 
    \[
    \Delta (\zz) \ge g(\zz) - g(\zz) + h(\xx, \yy) - h(\xx, \yy) = 0 \,, 
    \]
    and for all $\tilde \zz = (\tilde \xx, \tilde \yy) \in X \times Y$, we have 
    \[
    \begin{aligned}
    \Delta (\zz) &\ge \frac 1 2 \left[ g(\zz) - g(\xx, \tilde \yy) + h (\xx, \tilde \yy) - h(\xx, \yy) \right] + \frac 1 2 \left[ g(\zz) - g(\tilde \xx, \yy) + h (\xx, \yy) - h(\tilde \xx, \yy) \right] \\
    &= \frac 1 2 \left[ 2 g (\zz) + u_2 (\xx, \tilde \yy) + u_1 (\tilde \xx, \yy) \right] \\
    &= \frac 1 2 \left[ u_1 (\tilde \xx, \yy) - u_1 (\xx, \yy) + u_2 (\xx, \tilde \yy) - u_2 (\xx, \yy) \right] \,. 
    \end{aligned}
    \]
\end{proof}

\begin{proof}[Proof of \cref{proposition:delta root iff NE}]
    The (if) part follows directly from \cref{proposition:delta is the upper bound for NE approximation error}. Now we prove the (only if) part. Suppose $\zz^* = (\xx^*, \yy^*)$ is the Nash equilibrium. For all $\tilde \zz = (\tilde \xx, \tilde \yy) \in X \times Y$, 
    \[
    g (\zz ^*) - g (\tilde \zz) + h (\xx^*, \tilde \yy) - h (\tilde \xx, \yy^*) \le \lin {\nabla g(\zz ^*), \zz ^* - \tilde \zz} + \lin {\cH (\zz ^*), \zz ^* - \tilde \zz} \le 0 \,, 
    \]
    where in the first inequality we use \cref{assumption:convex-concave-combination,assumption:convex average}.
    Then, we have $\Delta (\zz ^*) = 0$. 
\end{proof}

\subsection{Proofs for the results in \texorpdfstring{\Cref{sec:convergence analysis}}{}}
\label{sec:proof of convergence}

The main technical work in the convergence analysis is to use the properties of our potential function and prove the descent lemma~(\cref{lemma:descent lemma}). 
\begin{proof}[Proof of \cref{lemma:descent lemma}]
    By \cref{assumption:convex-concave-combination}, we can upper bound the convex-concave zero-sum part
    \begin{equation}
    \label{eq:convex concave descent of h}
    \begin{aligned}
    h(\xx _{t+1}, \yy ^*) - h (\xx ^*, \yy _{t+1}) 
    &= h(\xx _{t+1}, \yy _{t+1}) - h (\xx ^*, \yy _{t+1}) + h(\xx _{t+1}, \yy ^*) - h(\xx _{t+1}, \yy _{t+1}) \\ 
    &\le \lin {\nabla _\xx h (\xx_{t+1}, \yy_{t+1}), \xx_{t+1} - \xx^*} - \frac {\mu} {2} \norm {\xx _{t+1} - \xx ^*} ^2 \\
    &\qquad - \lin {\nabla _\yy h (\xx_{t+1}, \yy_{t+1}), \yy_{t+1} - \yy^*} - \frac {\nu} {2} \norm {\yy _{t+1} - \yy ^*} ^2 \\
    &= \lin {\cH (\zz _{t+1}), \zz _{t+1} - \zz ^*} - \frac {\mu} {2} \norm {\xx _{t+1} - \xx ^*} ^2 - \frac {\nu} {2} \norm {\yy _{t+1} - \yy ^*} ^2 \,. 
    \end{aligned}
    \end{equation} 
    By \cref{assumption:convex average,assumption:delta-near-zero-sum}, we can upper bound the jointly convex coupling part 
    \begin{equation}
    \label{eq:smooth conevx three points descent lemma of g}
    \begin{aligned}
    g (\zz _{t+1}) - g (\zz ^*) 
    &= g(\zz _{t+1}) - g(\zz_t) + g(\zz _t) - g(\zz ^*) \\
    &\le \lin {\nabla g(\zz _t), \zz _{t+1} - \zz_t} + \frac \delta 2 \norm {\zz _{t+1} - \zz _t} ^2 + \lin {\nabla g(\zz _t), \zz _{t} - \zz ^*} \\
    &= \lin {\nabla g(\zz _t), \zz _{t+1} - \zz ^*} + \frac \delta 2 \norm {\zz _{t+1} - \zz _t} ^2 \,. 
    \end{aligned}
    \end{equation}
    
    In view of
    \begin{equation}
    \label{eq:cosine law}
    \lin { \zz _{t+1} - \zz _{t} , \zz _{t+1} - \zz ^*} = \frac 1 2 \norm {\zz _{t+1} - \zz ^*} ^2 - \frac 1 2 \norm {\zz _{t} - \zz ^*} ^2 + \frac 1 2 \norm {\zz _{t+1} - \zz _{t}} ^2 \,, 
    \end{equation}
    and 
    \begin{equation}
    \label{eq:subproblem approximation bound}
    \begin{aligned}
    &\quad \lin {\nabla g (\zz_t) + \cH (\zz _{t+1}) + \frac 1 {\eta_t} (\zz _{t+1} - \zz _{t}), \zz _{t+1} - \zz ^*} \\
    &\le \lin {\nabla _{\xx} \phi_t (\zz _{t+1}), \xx _{t+1} - \xx^*} - \lin {\nabla _{\yy} \phi_t (\zz _{t+1}), \yy _{t+1} - \yy^*} \\
    &\refLE{eq:subproblem accuracy} \epsilon_t \,,
    \end{aligned}
    \end{equation}
    we have 
    \[
    \begin{aligned}
    0 &= - \Delta (\zz ^*) \le g (\zz _{t+1}) - g (\zz ^*) + h(\xx _{t+1}, \yy ^*) - h (\xx ^*, \yy _{t+1}) \\
    &\stackrel{\eqref{eq:convex concave descent of h}\eqref{eq:smooth conevx three points descent lemma of g}}{\le} \lin {\nabla g(\zz_t) + \cH (\zz _{t+1}) , \zz_{t+1} - \zz ^*} - \frac {\mu} {2} \norm {\xx_{t+1} - \xx ^*} ^2 - \frac {\nu} {2} \norm {\yy_{t+1} - \yy ^*} ^2 + \frac \delta 2 \norm {\zz _{t+1} - \zz _{t}} ^2 \\
    &= \lin {\nabla g(\zz_t) + \cH (\zz _{t+1}) + \frac 1 {\eta_t} (\zz _{t+1} - \zz _{t}), \zz_{t+1} - \zz ^*} - \frac 1 {\eta_t} \lin {\zz _{t+1} - \zz _{t}, \zz_{t+1} - \zz ^*} \\
    &\qquad - \frac {\mu} {2} \norm {\xx_{t+1} - \xx ^*} ^2 - \frac {\nu} {2} \norm {\yy_{t+1} - \yy ^*} ^2 + \frac \delta 2 \norm {\zz _{t+1} - \zz _{t}} ^2 \\
    &\stackrel{\eqref{eq:cosine law}\eqref{eq:subproblem approximation bound}}{\le} \epsilon_t + \frac 1 {2 \eta_t} \norm {\xx_t - \xx ^*} ^2 - \left( \frac 1 {2 \eta_t} + \frac {\mu} {2} \right) \norm {\xx_{t+1} - \xx ^*} ^2 \\
    &\qquad + \frac 1 {2 \eta_t} \norm {\yy_t - \yy ^*} ^2 - \left( \frac 1 {2 \eta_t} + \frac {\nu} {2} \right) \norm {\yy_{t+1} - \yy ^*} ^2 - \left( \frac 1 {2 \eta_t} - \frac {\delta} {2} \right) \norm {\zz _{t+1} - \zz _{t}} ^2 \,,
    \end{aligned}
    \]
    where the first equality follows from \cref{proposition:delta root iff NE} and the first inequality follows from the definition of $\Delta (\cdot)$. 
    Finally, the desired bound follows from $\eta _t \le \frac 1 {\delta}$. 
\end{proof}

With \cref{lemma:descent lemma}, we are ready to prove the complexity of the outer loop~(\cref{lemma:outer loop}). 
\begin{proof}[Proof of \cref{lemma:outer loop}]
    For monotone $\delta$-nearly-zero-sum games and $\eta \le \frac 1 \delta$, by \cref{lemma:descent lemma}, for any $k \in [0, t-1] \cap \mathbb Z$, we have  
    \[
    \norm {\zz _{k+1} - \zz ^*} ^2 \le (1-\theta) \norm {\zz _{k} - \zz ^*} ^2 + 2 \eta \epsilon_k \,. 
    \]
    Then, unrolling this recursion (from $k = t-1$, $t-2$, $\cdots$, to $0$) yields
    \[
    \begin{aligned}
    \norm {\zz _{t} - \zz ^*} ^2 
    &\le (1-\theta) ^t \norm{\zz _0 - \zz^*} ^2 + 2 \eta \sum _{k=0} ^{t-1} (1-\theta) ^{t-k-1} \epsilon _{k} \\
    &\le (1-\theta) ^t (D_X^2 + D_Y^2) + \frac {2 \eta} \theta \cdot \max _{k \in [0,t-1] \cap \mathbb Z} \epsilon _{k} \\
    &\le \frac{\epsilon}{2} + \frac{\epsilon}{2} \\
    &= \epsilon \,,
    \end{aligned}
    \]
    where the last inequality follows from \( 
    t \ge \frac 1 {\theta} \log \frac {2 (D_X^2 + D_Y^2)} {\epsilon} 
    \) and $\epsilon _t \le { \frac {\theta \epsilon} {4 \eta}}$. 
\end{proof}

Below, we also include the proof of the gradient complexity of the inner loops for completeness. This result of the inner loops is heavily based on the previous results of optimal gradient methods in minimax optimization (see, for instance, \cite{kovalev2022first,carmon2022recapp,thekumparampil2022lifted,lan2023novel}). 
\begin{proof}[Proof of \cref{lemma:inner loop}]
    Let $\zz _{t+1} ^* = (\xx _{t+1} ^*, \yy _{t+1} ^*) \in X \times Y$ denote the saddle point of $\phi _t (\cdot, \cdot)$. 
    Denote \[ \bar \epsilon_t = \frac{\epsilon_t^2} {8 L^2 \left( D_X^2 + D_Y^2 \right)} \,. \] 
    By \cref{proposition:approximate-gradient-distance}, an inexact solution in \cref{eq:subproblem accuracy} of \cref{algorithm:ICL} can be obtained from a pair of decisions $\bar \zz_{t+1} = (\bar \xx_{t+1}, \bar \yy_{t+1}) \in X \times Y$ that satisfies $\norm {\bar \zz_{t+1} - \zz_{t+1} ^*}^2 \le \bar \epsilon_t$. 

    The function $\phi_t (\cdot, \cdot)$ is $\left( \eta_t^{-1} + \mu \right)$-strongly convex-$\left( \eta_t^{-1} + \nu \right)$-strongly concave and $2L$-smooth, where the $2L$-smoothness follows from $\eta _t \ge \frac 1 L$. Hence, by \cite{kovalev2022first,carmon2022recapp,thekumparampil2022lifted,lan2023novel}, the aforementioned pair of decisions $\bar \zz_{t+1}$ can be found within 
    \[
    \cO \left( \frac L {\sqrt {\left(\eta_t ^{-1} + \mu\right) \left(\eta_t ^{-1} + \nu\right)}} \cdot \log \left( \frac {D_X^2 + D_Y^2} { \bar \epsilon_t } \right) \right) 
    \]
    gradient queries. Finally, after substituting the $\bar \epsilon_t$, the desired bound follows. 
\end{proof}

Finally, we prove \cref{thm:total gradient queries}, our main theoretical result. 

\begin{proof}[Proof of \cref{thm:total gradient queries}]
    The overall gradient complexity is given by the multiplication of outer loop iterations~(\cref{lemma:outer loop}) and inner loop gradient complexity~(\cref{lemma:inner loop}): 
    \begin{small}
    \[
    \begin{aligned}
        &\quad \cO \left( \frac {\eta^{-1} + \min \{\mu, \nu\}} {\min \{\mu, \nu\}} \cdot \log \frac {2 (D_X^2 + D_Y^2)} {\epsilon} \right) \cdot \cO \left( \frac L {\sqrt {\left(\eta ^{-1} + \mu\right) \left(\eta ^{-1} + \nu\right)}} \cdot \log \frac {L (D_X^2 + D_Y^2)} { \epsilon_t } \right) \\
        &= \cO \left( \frac {\delta + \min \{\mu, \nu\}} {\min \{\mu, \nu\}} \cdot \log \frac {D_X^2 + D_Y^2} {\epsilon} \right) \cdot \cO \left( \frac L {\sqrt {\left(\delta + \mu\right) \left(\delta + \nu\right)}} \cdot \log \left( \frac {L (D_X^2 + D_Y^2)} { \min \{\mu, \nu\} \cdot \epsilon } \right) \right) \\
        &= \cO \left( \frac {L} {\min \{\mu, \nu\}} \cdot \sqrt{ \frac {\delta + \min \{\mu, \nu\}} {\delta + \max \{\mu, \nu\}} } \cdot \log \left( \frac { L \left(D_X^2 + D_Y^2\right) } {\min \{\mu,\nu\} \cdot \epsilon} \right) \log \left( \frac{D_X^2+D_Y^2}{\epsilon} \right) \right) \\
        &= \cO \left( \left(\frac {L} {\sqrt {\mu \nu}} + \frac {L} {\min \{\mu, \nu\}} \cdot \min \left\{ 1, \sqrt{ \frac {\delta} {\mu + \nu} } \right\} \right) \cdot \log \left( \frac { L \left(D_X^2 + D_Y^2\right) } {\min \{\mu,\nu\} \cdot \epsilon} \right) \log \left( \frac{D_X^2+D_Y^2}{\epsilon} \right) \right) \,, 
    \end{aligned}
    \]
    \end{small}
    where the first relation follows from $\eta = \min \left\{ \frac 1 \delta, \frac 1 {\min \{\mu, \nu\}} \right\}$. 
\end{proof}

\subsection{Proofs for the result in \texorpdfstring{\Cref{sec:non-strongly monotone}}{}}
\label{sec:proof of non-strongly monotone}

We state a more general formulation of \cref{corollary:our informal non-strongly monotone rates}. 
\begin{corollary}
\label{corollary:our non-strongly monotone rates}
    For monotone $\delta$-near-zero-sum games where $\mu=0$ or $\nu=0$, an $\epsilon$-approximate Nash equilibrium can be found within 
    \[
    \cO \left( \left(\frac {L} {\sqrt {\bar \mu \bar \nu}} + \frac {L} {\min \{\bar \mu, \bar \nu\}} \cdot \min \left\{ 1, \sqrt{ \frac {\delta} {\bar \mu + \bar \nu} } \right\} \right) \cdot \log \left( \frac { L^2 {D^2} } {\min \{\bar \mu,\bar \nu\} \cdot \epsilon} \right) \log \left( \frac{L {D^2} }{\epsilon} \right) \right) 
    \]
    gradient queries, where $\bar \mu = \mu + \min \left\{ \frac{\epsilon}{2 D_X^2}, L \right\}$ and $\bar \nu = \nu + \min \left\{ \frac{\epsilon}{2D_Y^2}, L \right\}$. 
\end{corollary}

\begin{proof}[Proof of \cref{corollary:our non-strongly monotone rates}]
    We consider the reduced game where \pone\ maximizes \[ \hat u_1 = u_1 - \min \left\{ \frac{\epsilon}{4 D_X^2}, \frac L 2 \right\} \norm {\xx}^2 + \min \left\{ \frac{\epsilon}{4 D_Y^2}, \frac L 2 \right\} \norm{\yy}^2 \] over $\xx \in X$ and \ptwo\ maximizes \[ \hat u_2 = u_2 + \min \left\{ \frac{\epsilon}{4 D_X^2}, \frac L 2 \right\} \norm {\xx}^2 - \min \left\{ \frac{\epsilon}{4 D_Y^2}, \frac L 2 \right\} \norm{\yy}^2\] over $\yy \in Y$. Any $\frac \epsilon 2$-approximate Nash equilibrium of the reduced game is an $\epsilon$-approximate Nash equilibrium in the original game. This reduction is similar to the ones used in \cite{lin2020near,wang2020improved}. 
    
    Denote $\hat g \triangleq - \frac 1 2 (\hat u_1 + \hat u_2)$ and $\hat h \triangleq \frac 1 2 (-\hat u_1 + u_2)$. Then, we have $\hat h = h + \left\{ \frac{\epsilon}{4 D_X^2}, \frac L 2 \right\} \norm {\xx}^2 - \left\{ \frac{\epsilon}{4 D_Y^2}, \frac L 2 \right\} \norm {\yy}^2$, which is $2L$-smooth and $\bar \mu$-strongly convex-$\bar \nu$-strongly concave. We also have $\hat g = - \frac 1 2 (\hat u_1 + \hat u_2) = - \frac 1 2 (u_1 + u_2) = g$, which is jointly convex and $\delta$-smooth. By \cref{thm:total gradient queries}, we obtain the number of gradient queries for an $\frac {\epsilon^2} {32 L^2 D^2}$-accurate Nash equilibrium in the reduced game: 
    \[
    \cO \left( \left(\frac {L} {\sqrt {\bar \mu \bar \nu}} + \frac {L} {\min \{\bar \mu, \bar \nu\}} \cdot \min \left\{ 1, \sqrt{ \frac {\delta} {\bar \mu + \bar \nu} } \right\} \right) \cdot \log \left( \frac { L^2 {D^2} } {\min \{\bar \mu,\bar \nu\} \cdot \epsilon} \right) \log \left( \frac{L {D^2} }{\epsilon} \right) \right) \,. 
    \]

    Finally, following from \cref{proposition:approximate-gradient-distance}, we obtain the desired $\frac \epsilon 2$-approximate Nash equilibrium of the reduced game by taking an extragradient step from the $\frac {\epsilon^2} {32 L^2 D^2}$-accurate Nash equilibrium. 
\end{proof}

\subsection{Proofs for the results in \texorpdfstring{\Cref{sec:application examples}}{}}
\label{sec:proof of examples}

\begin{proposition}[Convex reformulation in bilinear coupling]
    For $\beta_1, \beta_2\ge 0$ and ${\mM} \in R^{m \times n}$ such that $\sqrt{\beta_1 \beta_2} \ge \norm {\mM}$, the function $\tilde g(\cdot,\cdot): \R^n \times \R^m \to \R$ defined as 
    \[
    \tilde g (\xx, \yy) = \frac {\beta_1} 2 \norm {\xx}^2 + \lin {\mM \xx, \yy} + \frac {\beta_2} 2 \norm {\yy}^2  
    \]
    is jointly convex. 
\end{proposition}
\begin{proof}
    The quadratic function $\tilde g(\cdot, \cdot)$ is bounded below: for all $\xx \in \R^n$, $\yy \in \R^m$, 
    \[
    \begin{aligned}
    \tilde g (\xx, \yy)
    &\ge \frac {\beta_1} {2} \norm{\xx}^2 - \norm {\mM \xx} \norm {\yy} + \frac {\beta_2} {2} \norm{\yy}^2 \\ 
    &\ge \frac {\beta_1} {2} \norm{\xx}^2 - \sqrt{\beta_1 \beta_2} \norm {\xx} \norm {\yy} + \frac {\beta_2} {2} \norm{\yy}^2 \\ 
    &\ge 0 \,, 
    \end{aligned}
    \]
    where in the first inequality we used the Cauchy-Schwarz inequality. Therefore, $\tilde g (\cdot, \cdot)$ is jointly convex. 
\end{proof}

\begin{proposition}[Convex reformulation in general coupling]
    For $\beta \ge 0$ and ${g} : X \times Y \rightarrow \R$ such that $g(\cdot,\cdot)$ is $\beta$-smooth, the function $\tilde g(\cdot,\cdot): X \times Y \to \R$ defined as 
    \[
    \tilde g(\xx, \yy) = \frac {\beta} 2 \norm {\xx}^2 + g (\xx, \yy) + \frac {\beta} 2 \norm {\yy}^2  
    \]
    is jointly convex. 
\end{proposition}
\begin{proof} 
    For all $\zz= (\xx, \yy) \in X \times Y$ and $\zz^\prime = (\xx^\prime, \yy^\prime) \in X \times Y$, we have 
    \[
    \begin{aligned}
    \lin {\nabla \tilde g (\zz^\prime) - \nabla \tilde g(\zz), \zz^\prime - \zz} 
    &= \beta \norm {\zz^\prime - \zz}^2 + \lin {\nabla g (\zz^\prime) - \nabla g(\zz), \zz^\prime - \zz} \\
    &\ge \beta \norm {\zz^\prime - \zz}^2 - \beta \norm {\zz^\prime - \zz}^2 \\
    &= 0 \,,
    \end{aligned}
    \]
    where the first inequality follows from the $\beta$-smoothness of $g(\cdot, \cdot)$. Therefore, the function $\tilde g(\cdot, \cdot)$ is jointly convex~\citep[Theorem~2.1.3]{nesterov2004introductory}. 
\end{proof}

\section{Additional results for other oracle and function classes}
\label{sec:additional results for other oracle and function classes}

In this section, we consider a different class of Nash equilibrium problem and demonstrate the applicability of our \algname{ICL} framework. In \cite{boct2023accelerated}, they considered a zero-sum (or strictly competitive) game where the two players have proximal oracle and gradient oracle, respectively. We now consider the generalization where an additional incentive is added. To prevent ambiguity, we will define the problem class in a self-contained way. 

In this section, we are interested in the Nash equilibrium problem, or equivalently, the variational inequality problem given by operator $\cF$ defined on $X \times Y$: 
\[
\cF (\xx, \yy) = \nabla g (\xx, \yy) + \left( \nabla _\xx h (\xx, \yy), - \nabla _\yy h (\xx, \yy) + \nabla \psi (\yy) \right), \quad (\xx, \yy) \in X \times Y , 
\]
where 
\begin{enumerate}
    \item The sets $X$ and $Y$ are compact convex sets in Euclidean spaces. The diameter of $X$ is bounded by $D_X$, and the diameter of $Y$ is bounded by $D_Y$. 
    \item The function $g \colon X \times Y \to \R$ is $\delta$-smooth and convex. 
    \item The function $\psi \colon Y \to \R \cup \{ +\infty \}$ is proper, lower semicontinuous, $\nu$-strongly convex, and with domain 
    \(
    \dom \psi = \left\{ \yy \in Y \mid \psi(y) < +\infty \right\} 
    \). 
    \item For all $\yy \in \dom \psi$, the function $h (\cdot, \yy) \colon X \to \R \cup \{ + \infty \}$ is proper, lower semi-continuous, and $\mu$-strongly convex. 
    \item For all $\xx \in \Pi _{X} (\dom h) \triangleq \left\{ \uu \in X \mid \exists \vv \in Y \text{ such that } (\uu, \vv) \in \dom h \right\}$, we have that $\dom h (\xx, \cdot) = Y$ and the function $h (\xx, \cdot) \colon Y \to \R$ is concave and continuously differentiable. Moreover, $\Pi _{X} (\dom h)$ is closed. 
    \item There exists $L_{yx}, L_{yy} \ge 0$ such that for all $(x, y), (x^\prime, y^\prime) \in \Pi _X \left( \dom h \right) \times \dom \psi$, 
    \[
    \norm {\nabla _\yy h (\xx, \yy) - \nabla _\yy h (\xx^\prime, \yy^\prime)} \le L_{yx} \norm {\xx - \xx^\prime} + L_{yy} \norm {\yy - \yy^\prime} .
    \]
\end{enumerate}
We assume the players can query the gradient $\nabla g$, the proximal operator of $h(\cdot,\yy)$ for any fixed $\yy \in Y$, the partial gradient $\nabla _\yy h (\cdot, \cdot)$, and the proximal oracle of $\psi (\cdot)$. 
The problem studied in \cite{boct2023accelerated} corresponds to a special case of the additional incentive $\delta = 0$, while we generalize their results to $\delta \neq 0$. 

We will use the complexity results in \cite{boct2023accelerated} as a black box. We cite their results below. 

\begin{lemma}[\cite{boct2023accelerated}, Theorem~14]
\label{lemma:grad prox complexity}
    For $\delta=0$ and $\mu > 0$, there exists an algorithm which returns an $\epsilon$-accurate Nash equilibrium with the number of partial gradient queries to $\nabla _\yy h (\cdot)$ and the number of proximal oracle queries to $\prox _{h(\cdot, \yy)} (\cdot)$ and $\prox _{\psi} (\cdot)$ bounded by 
    \[
    \cO \left( \left( 1 + \frac{L_{yx}}{\sqrt{\mu \nu}} + \frac{L_{yy}}{\nu} \right) \cdot \log \left(\frac {D^2} \epsilon\right) \right) \,. 
    \]
\end{lemma}

By applying our \algname{ICL} algorithm~(\cref{algorithm:ICL}), we obtain the following complexity result: 
\begin{theorem}
    Assume $h$ is $L$-smooth over $\Pi _X \left( \dom h \right) \times \dom \psi$, and $\psi$ is $L$-smooth over $\dom \psi$. For $\mu > 0$, there exists an algorithm which returns an $\epsilon$-accurate Nash equilibrium with the number of partial gradient queries to $\nabla _\yy h (\cdot)$ and the number of proximal oracle queries to $\prox _{h(\cdot, \yy)} (\cdot)$ and $\prox _{\psi} (\cdot)$ bounded by 
    \[
    \cO \left( \left( 1 + \frac{\delta}{\min \{\mu, \nu\}} \right) \left( 1 + \frac{L_{yx}}{\sqrt{ \left(\delta + \mu \right) \left( \delta + \nu \right)} } + \frac{L_{yy}}{\delta + \nu} \right) \cdot \log \left( \frac {LD^2}{\min\{\mu, \nu\} \cdot \epsilon} \right) \log \left(\frac {D^2} \epsilon\right) \right) \,, 
    \]
    and with the number of gradient queries to $\nabla g (\cdot)$ bounded by 
    \[
    \cO \left( \left( 1 + \frac{\delta}{\min \{\mu, \nu\}} \right) \cdot \log \left(\frac {D^2} \epsilon\right) \right) \,. 
    \]
\end{theorem}

\begin{proof}
    The result follows from multiplying the outer loop iterations in \cref{lemma:outer loop} and the inner loop complexity in \cref{lemma:grad prox complexity}. 
\end{proof}

% \itodo{Start from the first two problems. Should be easier.}

% \subsection{Convex-strongly concave zero-sum part with linear minimization oracles}

% \cite{chen2020efficient} is not good. We need to reduce to their strongly convex-strongly concave results.

% \subsection{Strongly convex-strongly concave zero-sum part with gradient oracle and proximal oracle}

% \subsection{P\L -P\L\ zero-sum part with gradient oracles}

% \cite{chen2022faster}

% \itodo{This might be the most difficult one. Start from proximal point methods for P\L\ functions. \url{https://optimization-online.org/wp-content/uploads/2016/08/5590.pdf} Need a new descent lemma with P\L\ functions}
\section{Illustration of the application examples}
\label{sec:illustration}

\paragraph{Matrix games with transaction fees}
We give a concrete illustration for matrix games with transaction fees. Let the payoff matrices of \pone\ and \ptwo\ without transaction fees be
\[
\mM = 
\begin{bmatrix}
    300 & -200 \\
    -100 & 400
\end{bmatrix}
\quad \text{and} \quad 
-\mM = 
\begin{bmatrix}
    -300 & 200 \\
    100 & -400
\end{bmatrix}
\,,
\]
respectively. Then, 
\[
\abs{\mM} = \begin{bmatrix}
    300 & 200 \\
    100 & 400
\end{bmatrix}, \quad
\mM_+ = \begin{bmatrix}
    300 & 0 \\
    0 & 400
\end{bmatrix}, \quad \text{and} \quad
\mM_- = \begin{bmatrix}
    0 & 200 \\
    100 & 0
\end{bmatrix} \,. 
\]

Let $1\%$ of transaction fees be imposed on every payment. Then, the payoff matrices of \pone\ and \ptwo\ with transaction fees are 
\[
\mA = 
\begin{bmatrix}
    297 & -200 \\
    -100 & 396
\end{bmatrix}
\quad \text{and} \quad 
\mB = 
\begin{bmatrix}
    -300 & 198 \\
    99 & -400
\end{bmatrix}
\,,
\]
respectively. We also draw the \cref{tab:pre-transaction and post-transaction} for easier comparisons. 

\begin{table}[h!]
\centering
\caption{An illustration of matrix games with transaction fee $\rho = 0.01$.}
\label{tab:pre-transaction and post-transaction}
\begin{tabular}{|c|c|}
\hline
300/-300 & -200/200 \\
\hline
-100/100 & 400/-400 \\
\hline
\end{tabular}
\hspace{1cm}
\begin{tabular}{|c|c|}
\hline
297/-300 & -200/198 \\
\hline
-100/99 & 396/-400 \\
\hline
\end{tabular}
\end{table}

\blue{
\paragraph{Competitive games with small cooperation incentives}
We give a concrete illustration for competitive games with small cooperation incentives. The donation games are a canonical model in evolutionary game theory for studying the altruistic collaboration~\citep{nowak2006evolutionary,sigmund2010calculus}. 

Let us consider a simplified version of donation games played for one round, as shown in \cref{tab:donate game}. 
In this game, a player can choose to incur a personal cost, \(c\), to provide a larger benefit, \(b\), to another player. Consider a concrete example where the cost to donate is \(c=50\) and the benefit conferred is only slightly higher at \(b=51\). 

In this scenario, if both players cooperate, they each pay a cost of $50$ and receive a benefit of $51$, resulting in a modest net payoff of $1$. However, the temptation to defect is substantial: by withholding their own donation while still receiving the benefit from the other player, a defector can achieve a payoff of $51$. Conversely, the cooperating player who is defected upon is left with a ``sucker's payoff'' of $-50$. This low benefit-to-cost ratio (\(b/c \approx 1.02\)) creates a very small incentive for cooperation. 

\begin{table}[h!]
\centering
\blue{
\caption{An illustration of donation games.}
}
\label{tab:donate game}
\blue{
\begin{tabular}{l|c|c|}
  \multicolumn{1}{c}{} & \multicolumn{1}{c}{Player 2: Cooperate} & \multicolumn{1}{c}{Player 2: Defect} \\
  \cline{2-3}
  Player 1: Cooperate & (1, 1)  & (-50, 51) \\
  \cline{2-3}
  Player 1: Defect    & (51, -50) & (0, 0)  \\
  \cline{2-3}
\end{tabular}
}
\end{table}
}

\section{More experimental details}
\label{sec:more details and additional experiments}

\subsection{Implementation details}

We generate the sparse matrix $\mM$ following the procedures outlined in \cite{nemirovski2004prox,nesterov2005smooth}: (i) The random seeds are set from $0$, $111$, $222$, ..., and $999$; (ii) $100000$ coordinates of $\mM$ are chosen uniformly at random; (iii) each chosen coordinate is assigned a random value independently drawn from a uniform distribution between $[-1,1]$; (iv) all remaining coordinates are set to $0$. 

We implement our \algname{ICL} method as described in \cref{algorithm:ICL}.  
The classic \algname{OGDA} and classic \algname{EG} methods are implemented as outlined in \cite{popov1980modification} and \cite{korpelevich1976extragradient}, respectively. 
All solvers are initialized at $(\xx_0, \yy_0) = \left( \1_n /n, \1_m/m \right)$, where {$\1_k \in \R^k$ denotes the vector of size $k$ where every element in the vector is equal to $1$.} 
The setup for \algname{ICL} is detailed in \cref{thm:total gradient queries}. 
The stepsize for \algname{OGDA} is set to $\frac 1 {2L}$ following \cite{popov1980modification,mokhtari2020convergence}, and for \algname{EG} is set to $\frac 1 {\sqrt 2 L}$ following \cite{korpelevich1976extragradient,nemirovski2004prox}. 
% For the outer loop of \algname{ICL}, for \algname{OGDA}, and for \algname{EG}, we tune the stepsizes between $1\times$, $2\times$, $3\times$, $4\times$, and $5\times$ of the theoretical stepsize. 
For the inner loop, the Lifted Primal Dual method~\citep{thekumparampil2022lifted} is used, 
% To ensure fairness and consistent tuning efforts, 
with the theoretical setup maintained as specified in \cite[Theorem~2]{thekumparampil2022lifted}. 
% For all methods, we use the extra-gradient step described in \cref{proposition:extra gradient stopping criterion} as the stopping criterion. 

\subsection{More details of the experiment runs}

We conducted our experiments on \texttt{e2-highcpu} vCPUs within the \texttt{Google Cloud} environment. The memory requirement of our experiments is quite modest, requiring only sufficient RAM for a few $10000 \times 10000$ sparse matrices (that is, about $60$~MB). Each independent run completes within about $3$~minutes. 

\begin{sidewaysfigure}[p] % Use sidewaysfigure for floating figures; 'p' for a separate page
    \centering
    \includegraphics[width=\linewidth]{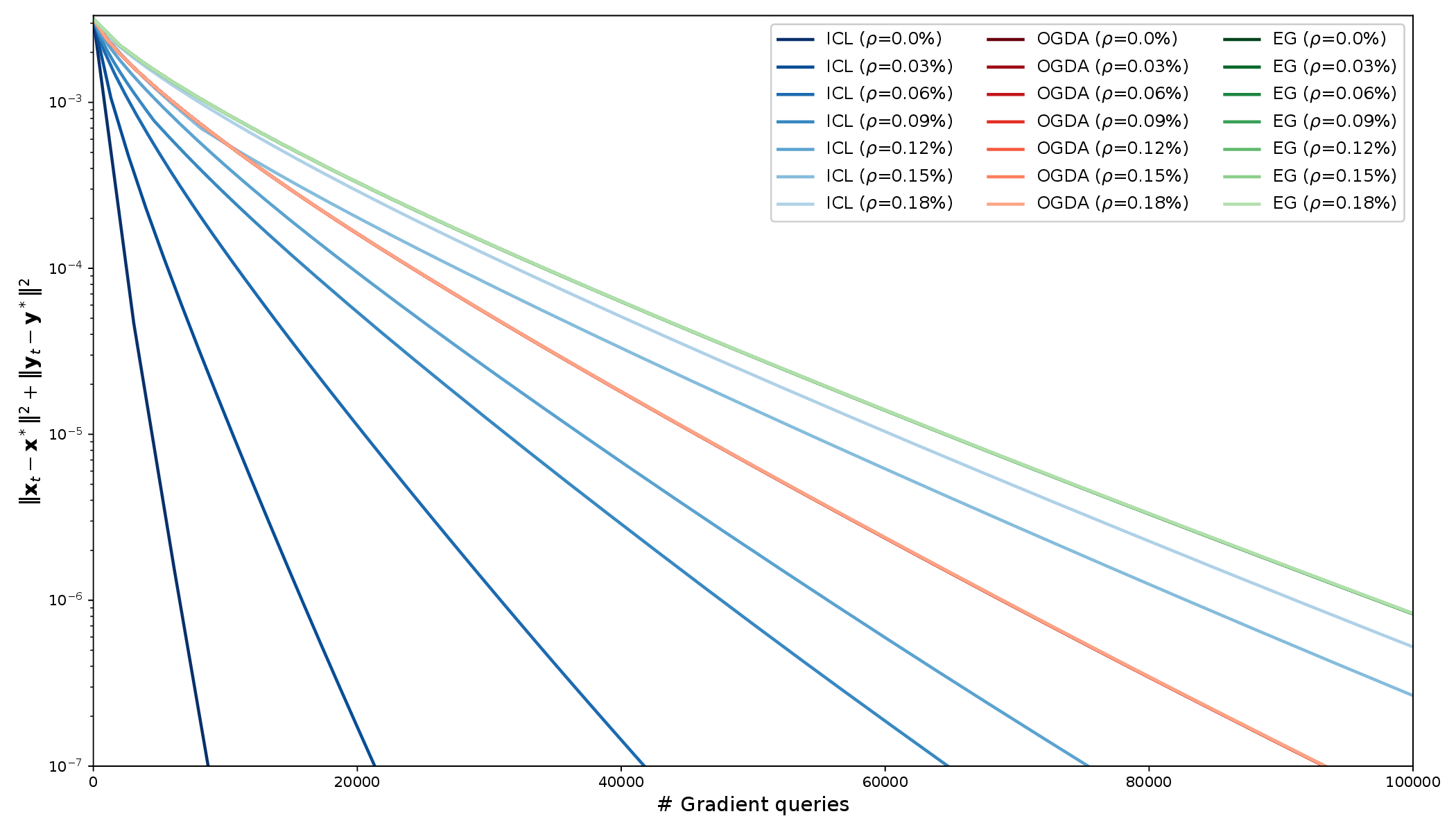}
    \caption{Comparisons of the convergence of the \algname{ICL}, \algname{OGDA}, and \algname{EG} methods with respect to the gradient query counts.}
    \label{fig:convergence wrt gradient queries}
\end{sidewaysfigure}

We plot the convergence behaviors in \cref{fig:convergence wrt gradient queries}. Note that for \algname{ICL}, only iterates within the outer loop are plotted. \Cref{fig:convergence wrt gradient queries} shows results for a single seed (seed $0$), as plotting all seeds in a single figure would introduce excessive visual complexity due to the unaligned $x$-axis representing the counts of gradient queries in the outer loop. Nonetheless, we observed consistent convergence patterns across different seeds: (i) Transaction fee changes have little impact on the convergence of \algname{OGDA} and \algname{EG}, but significantly accelerate the convergence of \algname{ICL} as $\rho$ decreases; (ii) \algname{ICL} converges fastest when $\rho \le 0.12\%$; and (iii) \algname{OGDA} converges fastest when $\rho \ge 0.15\%$. 

\blue{
We should clarify that all 21 entries (7 entries for each of the three algorithms) are indeed plotted in \cref{fig:convergence wrt gradient queries}. The perceived “lack of distinct lines” is actually a crucial empirical result supporting our theoretical claims:

\begin{itemize}
    \item For \algname{OGDA} (Red) and \algname{EG} (Green): The seven lines corresponding to different parameter settings for OGDA and EG almost completely merge and overlap. This empirical observation is fully consistent with our theory, which states that the convergence rate of classical variational inequality methods does not benefit from the near-zero-sum structure. Their rates are determined solely by $\frac{L}{\min \{\mu, \nu \}}$, regardless of the $\delta$ parameter. 
    \item For the proposed \algname{ICL} algorithm (Blue): The seven lines are clearly separated, showing a distinct dependency on the $\delta$ parameter. This separation empirically validates our key claim that our algorithm successfully harnesses the near-zero-sum structure to achieve a faster convergence rate. 
\end{itemize}
}

Finally, we report CPU times of experiment runs to converge to an $\epsilon$-accurate Nash equilibrium in \cref{tab:cpu time}, with error bars indicating $2$-sigma variations across $10$ independent runs using randomly generated matrices. \Cref{tab:cpu time} shows that \algname{ICL} achieves the shortest CPU time when $\rho \le 0.12\%$, while \algname{OGDA} achieves the shortest CPU time when $\rho \ge 0.15\%$. 
% We note that for $\rho = 0.15\%$, \algname{ICL}'s CPU time is \emph{longer} than \algname{OGDA}'s, despite the fact that \algname{ICL} exhibits slightly faster convergence to Nash equilibrium with respect to the gradient query counts~(see \cref{tab:results under different transaction rates} or \cref{fig:convergence wrt gradient queries}). This discrepancy may be due to \algname{ICL}'s nested loop structure resulting in higher CPU overhead. 
% ; (ii) Some inconsistencies between convergence to Nash equilibrium and satisfaction of the stopping criterion in \cref{proposition:extra gradient stopping criterion}. 

\begin{table}[ht]
    \centering
    \caption{The CPU times (in seconds) of the algorithms to converge to an $\epsilon$-accurate Nash equilibrium under various transaction fees. The error bars indicate $2$-sigma variations across the independent runs with $10$ randomly generated matrices.}
    \label{tab:cpu time}
    \resizebox{\linewidth}{!}{%
        \begin{tabular}{cccccccc}
          \toprule
          \diagbox[width=12em]{Methods}{Transaction fee $\rho$} & $0.00\%$ & $0.03\%$ & $0.06\%$ & $0.09\%$ & $0.12\%$ & $0.15\%$ & $0.18\%$ \\
          \midrule
          \algname{ICL}~(\cref{algorithm:ICL}) & $ \mathbf{19 \pm 0} $ & $ \mathbf{49 \pm 0} $ & $ \mathbf{93 \pm 0} $ & $ \mathbf{142 \pm 1} $ & $ \mathbf{167 \pm 0} $ & $ 247 \pm 2 $ & $ 264 \pm 3 $ \\
          \algname{OGDA}~\citep{popov1980modification} & $ 186 \pm 1 $ & $ 185 \pm 1 $ & $ 185 \pm 1 $ & $ 185 \pm 0 $ & $ 185 \pm 1 $ & $ \mathbf{185 \pm 1} $ & $ \mathbf{186 \pm 1} $ \\
          \algname{EG}~\citep{korpelevich1976extragradient} & $ 258 \pm 1 $ & $ 256 \pm 2 $ & $ 258 \pm 2 $ & $ 257 \pm 3 $ & $ 257 \pm 2 $ & $ 257 \pm 2 $ & $ 257 \pm 2 $ \\
          \bottomrule
        \end{tabular}
    }
\end{table}

\subsection{Additional runs under different parameter setting}
\label{subsec:additional runs}

In this section, we run additional numerical experiments under different parameter setting. We change the parameter $\nu = 0.01$, and we vary the transaction fee $\delta$ from $\{ 0.0\%, 0.3\%, \cdots, 1.8\% \}$. We keep the other parameter settings unchanged. 

The results, summarized in \cref{tab:results under different transaction rates more balanced higher transactiones}, demonstrate that \algname{ICL} requires fewer gradient queries to converge to an $\epsilon$-accurate Nash equilibrium when the transaction fee $\rho$ is below $1.2\%$. This empirical observation aligns with our theoretical prediction in \cref{example:transactioned matrix games}, which suggests that \algname{ICL} converges faster when $\rho \norm {\abs {\mM}} \lll \sqrt {\mu \nu} = 10 \%$. 

\begin{table}[ht]
\centering
\caption{Gradient query counts to converge to an $\epsilon$-accurate Nash equilibrium under various transaction fees. Error bars indicate $2$-sigma variations across the independent runs with $10$ randomly generated matrices.}
\label{tab:results under different transaction rates more balanced higher transactiones}
\resizebox{\linewidth}{!}{%
    \begin{tabular}{ccccccccc}
      \toprule
      \diagbox[width=12em]{Methods}{Transaction fee $\rho$} & $0.0\%$ & $0.3\%$ & $0.6\%$ & $0.9\%$ & $1.2\%$ & $1.5\%$ & $1.8\%$ \\
      \midrule
      \algname{ICL}~(\cref{algorithm:ICL}) & $ \mathbf{924 \pm 0} $ & $ \mathbf{924 \pm 0} $ & $ \mathbf{824 \pm 0} $ & $ \mathbf{1030 \pm 0} $ & $ \mathbf{1236 \pm 0} $ & $ 1648 \pm 0 $ & $ 2060 \pm 0 $ \\
      \algname{OGDA}~\citep{popov1980modification} & $ 1364 \pm 0 $ & $ 1364 \pm 0 $ & $ 1364 \pm 0 $ & $ 1361 \pm 4 $ & $ 1359 \pm 5 $ & $ \mathbf{1353 \pm 6} $ & $ \mathbf{1350 \pm 6} $ \\
      \algname{EG}~\citep{korpelevich1976extragradient} & $ 1848 \pm 0 $ & $ 1848 \pm 0 $ & $ 1848 \pm 0 $ & $ 1848 \pm 0 $ & $ 1848 \pm 0 $ & $ 1848 \pm 0 $ & $ 1848 \pm 0 $ \\
      \bottomrule
    \end{tabular}
}
\end{table}

We also observe that the CPU times in this setting are within $5$ seconds for all independent runs.

\end{document}